\documentclass{ws-rv9x6}
\usepackage{subfigure}   % required only when side-by-side / subfigures are used
\usepackage{ws-rv-thm}   % comment this line when `amsthm / theorem / ntheorem` package is used
\usepackage{ws-rv-van}   % numbered citation & references (default)
\usepackage{wrapfig,caption}
\usepackage{hyperref}

\makeindex
\newcommand{\taon}{$\tau$-lepton }
\newcommand{\taons}{$\tau$-leptons }
\newcommand{\Enu}{E$_\nu$}
\newcommand{\xmax}{\ensuremath{X_{\rm max}}}

\newcommand{\lsim}{\mathrel{\hbox{\rlap{\lower.75ex \hbox{$\sim$}} \kern-.3em \raise.4ex \hbox{$<$}}}}
\newcommand{\gsim}{\mathrel{\hbox{\rlap{\lower.75ex \hbox{$\sim$}} \kern-.3em \raise.4ex \hbox{$>$}}}}

\definecolor{pink1}{RGB}{226, 24, 166}

\begin{document}

\chapter[Space-based Ultra-High Energy Cosmic Ray Experiments]{Space-based Ultra-High Energy Cosmic Ray Experiments} %\label{SnC_chSn}

\author[John F. Krizmanic] {John F. Krizmanic\footnote{john.f.krizmanc@nasa.gov}}

\address{NASA/Goddard Space Flight Center\\
Laboratory for Astroparticle Physics  \\
%1000 Hilltop Circle \\
Greenbelt, Maryland 20771 USA}

%\begin{abstract}
\begin{center}
{\it In space, no one can hear you scream} $\cdots$
\newline {tagline of Ridley Scott's movie {\it Alien}} 
\end{center}
\begin{abstract}

Space-based experiments, either orbiting the Earth or from scientific balloon altitudes, measure high-energy cosmic 
rays by measuring from above the atmosphere the optical and radio signals generated by extensive air showers (EAS). These experiments are designed to have a large field-of-view (FoV) for observing EAS which translates to monitoring the atmosphere over a large, $\sim 10^6$ km$^2$ area on the ground.  Ultra-high energy cosmic rays (UHECRs, $E_{CR} \gsim$ 1 EeV) are measured by using the isotropic near-UV air fluorescence signal to finely sample the EAS development and to efficiently use the atmosphere as a vast calorimeter.  At UHE, these immense EAS particle cascades have sufficient charged particle content to generate the relatively dim fluorescence light that propagates to the space-based instrument. Additionally, the beamed Cherenkov light and geomagnetic radio emission from EAS arrive with $\gsim 10$ ns impulse and are measured at small angles away from the cosmic ray trajectories. In particular for optical Cherenkov measurements,  the energy thresholds can be $\gsim 1$ PeV, i.e. for very-high energy cosmic rays (VHECRs).  The instruments that use these EAS optical signals are effectively coarse-imaging telescopes with meter-sized optical collecting areas to minimize the VHECR and UHECR detection energy thresholds, which are in part set by the dark-sky airglow chemiluminescence optical background. Since optical signals need to be measured near astronomical night, these experiments have a mission-averaged measurement live time (duty cycle) around 10$-$20\%, which takes into effects of the variable viewing conditions, mainly cloud cover, high-altitude ozone thickness, and avoiding viewing areas with a high aurora backgrounds. The effects of low-altitude aerosols are small when viewing EAS via fluorescence from above the atmosphere, since the majority of the EAS development is above the $\sim 1$ km scale height of the aerosol layer. These effects are further complicated since the monitored atmospheric volume changes due to motion of the experiment. 

In this chapter, the nature of observing the UHECR-induced shower development from orbiting and balloon-borne experiments is detailed, both for missions that have been flown and those currently in development.  This will be accomplished by discussing experimental performance in terms of measuring the UHECR spectrum, UHECR nuclear composition, and UHECR arrival direction.  The ability for these experiments to also perform VHECR and VHE and UHE cosmic neutrino measurements will also be discussed. 
\end{abstract}

%\markboth{Even Page Header}{Odd Page Header} % Customized running heads

\body

%\tableofcontents

\section{Introduction}

\begin{figure}[h]
\begin{center}
    \includegraphics[width=0.99\textwidth]{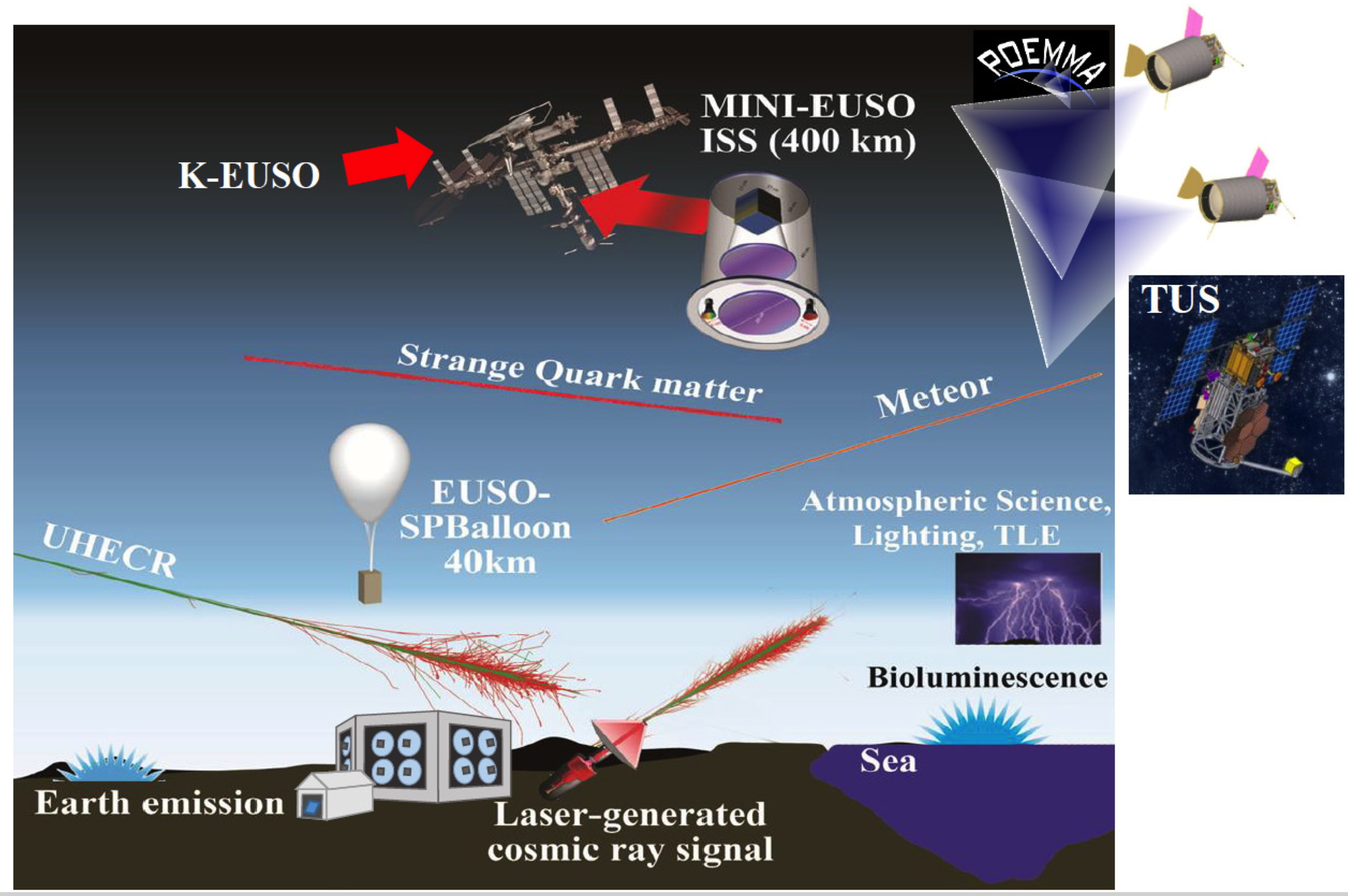}
 \end{center}
\vspace{-0.3cm}
  \caption{A composite collection of space-based experiments designed to measure UHECRS via the optical signals from extensive air showers (EAS), included several that have flown, TUS and Mini-EUSO; those that will fly in the near-term EUSO-SPB2; and those that are in the development stage, K-EUSO and POEMMA. UHECR measurements are the primary science goals but large-FoV that image with $0.1^\circ$ resolution in the near-UV and optical wavelengths are sensitive to the variety of secondary science signals, but with much longer observational time signatures than UHECRs. These include meteors, lightning and atmospheric transient luminous events, and bioluminescence. Also shown is the existing technique of using upward-pointed laser pulses to mimic EAS for instrument calibration. From Ref. \citenum{Schroeder:2019TB}, F. Schr\"{o}der: 2019 ICRC Rapporteur Presentation.
  DOI:\href{https://doi.org/10.22323/1.358.0030}{10.22323/1.358.0030}
  }
%\vspace{-3 mm}
  \label{UHECRcomposite}
\end{figure}

Space-based experiments use optical or radio telescopes that are designed to view from above the atmosphere the development of extensive air showers (EAS) initiated by ultra-high energy cosmic rays.  Since the EAS particle cascades have an immense content of charged particles, dominated by electrons and positrons even for hadronic UHECRS, they can form signals that can be detected by distant Earth orbiting or balloon-borne experiments.  The relatively course imaging requirements needed to accurately sample the $\gsim 10$ $\mu$s waxing and waning development using the EAS near-UV air fluorescence signal, $\sim$ km spatial scale near the Earth's surface, allows for the use of large field-of-view (FoV) telescopes and translates to monitoring $\sim 10^{13}$ tons of atmosphere for UHECRs. Fig.~\ref{UHECRcomposite} artistically presents the variety of EAS fluorescence and optical Cherenkov instruments that perform UHECR measurements as well the laser calibration technique and sources of other optical signals, including meteors and transient atmosphere phenomena \cite{Schroeder:2019TB}. and Since the majority of EAS development is within $\sim$ 30 km altitude above sea level, the large distance from a nominal 525 km low Earth orbit (LEO) to the EAS requires meter-scale optical systems, including refractive systems based on Fresnel lenses and reflective systems, such as Schmidt telescopes. However, even with these large optics, the threshold energy for UHECR detection is $\gsim 20$ EeV, which implies space-based UHECR experiments are designed to provide high exposure measurements of the highest energies of the cosmic ray spectrum.  The advantage of using an observations from LEO lead to full-sky sensitivity to cosmic UHECR sources on a time-scale of $\sim$ year due to the required operation near astronomical night to measure the optical EAS signal, which yields $\sim 10-15\%$ measurement average live time. These also yield all-sky measurement of cosmic UHECR sources under a single experimental framework with well-understood systematic errors on the measurements, especially when two satellites perform precise stereo fluorescence measurements that well determine the 3-dimensional location of key features of the EAS, including incident angle and the slant depth of EAS maximum, \xmax, which provides a measure of the identity of the UHECR particle. However, the volumes of atmosphere viewed by an orbiting experiment quickly change. Thus the UHECR measurements are performed in changing atmospheric conditions, in particular cloud cover in the FoV. Ancillary devices such as an IR camera measuring cloud top temperature and or LIDAR measurements to give more detailed information along the EAS viewing direction provide the needed information. This information will be augmented with that available from other Earth atmosphere monitoring satellites, such as used in the MERRA-2 global atmospheric properties database \cite{MERRA2}.

Experiments flying at scientific balloon altitudes, $\sim$ 30 km, also have sensitivity to UHECRs via air fluorescence measurements, but the UHECR exposure is limited due to the smaller atmospheric volume monitored by the telescope and due to shorter mission times.  Currently, the maximum float time of a balloon is at most $\sim$ 100 days assuming a ultra-long duration balloon (ULDB) flight. However, the ANITA long duration balloon (LDB) experiment has measured the beamed geomagnetic Cherenkov-like EAS radio emission from UHECRs both directly and reflected off the Antarctic ice for $E_{\rm CR} \gsim 1$ EeV \cite{2010PhRvL.105o1101H}. Additionally, simulation studies used to develop space-based experiments to detect optical Cherenkov light from EAS generated by cosmic neutrino interactions in the Earth have also shown that a meter-sized telescope pointed above the limb-of-the Earth will have sensitivity to VHECRs with $E_{\rm CR} \gsim 1$ PeV). While these measurements in the radio and optical Cherenkov are currently less precise than air fluorescence measurements, especially those done with EAS stereo reconstruction, they do offer the opportunity to measure cosmic rays at a much lower $\gsim$ PeV energy from space-based instruments.

\section{History}

\begin{figure}[ht]
\centerline{
\minifigure[Artist sketch of the space-based MASS mission concept. From Ref.~\citenum{MASS1996}.]
{\includegraphics[width=1.95in]{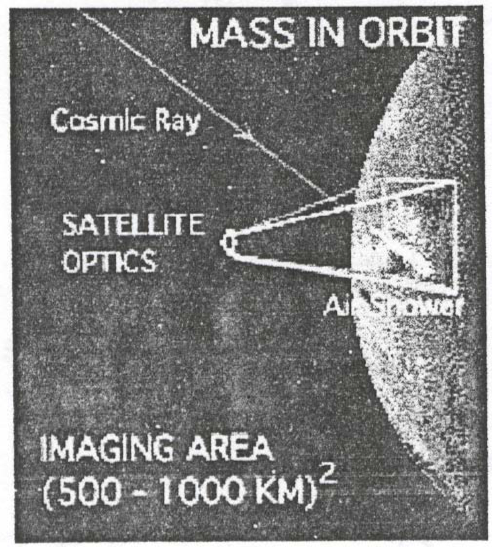}\label{MASS}}
\hspace{-8 mm}
\minifigure[Simulation graphic of two OWL spacecraft co-viewing UHECR-induced EAS, the field-of-views of the telescopes are highlighted. From Ref.~\citenum{2013ICRC...33.2334K}.]
{\includegraphics[width=2.3in]{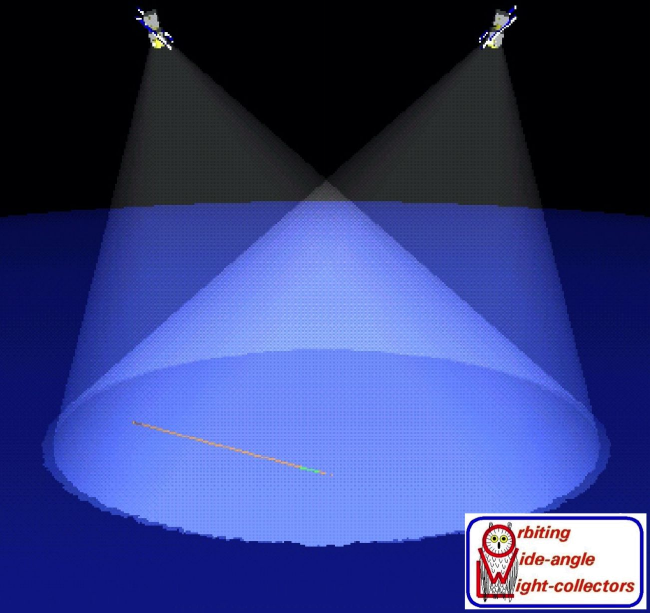}\label{OWLstereo}}
}
\vspace{-5mm}
\end{figure}

 The concept for observing ultra-high energy cosmic rays (UHECRs) from space was suggested in a paper by Robert Benson and John Linsley in 1980 \cite{1980BAAS...12Q.818B, 1981ICRC....8..145B}
\footnote{Folklore has it that the idea of observing  UHECR EAS from space was proposed during a during a coffee break conversation at a scientific workshop between John Linsley and other participants.}.
In a 1995 paper, Yoshi Takahasi proposed the concept called {\it Maximum-energy Auger (AIR)-Shower Satellite
(MASS)} \cite{1995ICRC....3..595T,1996SPIE.2806..102T}, to employ an Earth-orbiting,  wide Field-of-View (FoV) coarse-imaging telescope to view EAS development using the the near-UV fluorescence signal. A sketch of the MASS concept is shown in Fig.~\ref{MASS}, which was presented at a workshop on space-based UHECR cosmic ray measurements \cite{MASS1996}.

\begin{wrapfigure}{r}{0.5\columnwidth}
\captionsetup{width=0.46\columnwidth}
%\vspace{-4 mm}
\begin{center}
\includegraphics[width=0.48\textwidth]{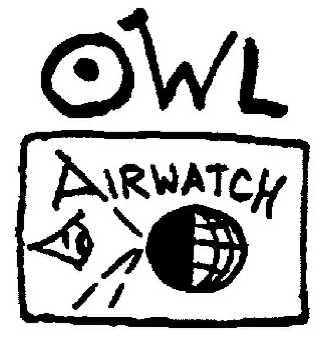}
\caption{The OWL/AIRWATCH logo drawn by John Linsley presented at an OWL/AIRWATCH meeting in 1998. \label{OWLairW}}
\end{center}
\vspace{-1mm}
\end{wrapfigure}

Important UHECR information is encoded in the development profile of an EAS: the total emitted optical light is proportional to the energy, sampling with sufficient temporal and spatial resolution the EAS evolution yields relatively long tracks to provide very good angular resolution for the UHECR primary, and measuring the location of shower maximum point in the development, \xmax, provides information about the primary, e.g. proton, nucleus, photon, or neutrino. 

\begin{figure}[ht]
\begin{center}
    \includegraphics[width=0.38\textwidth]{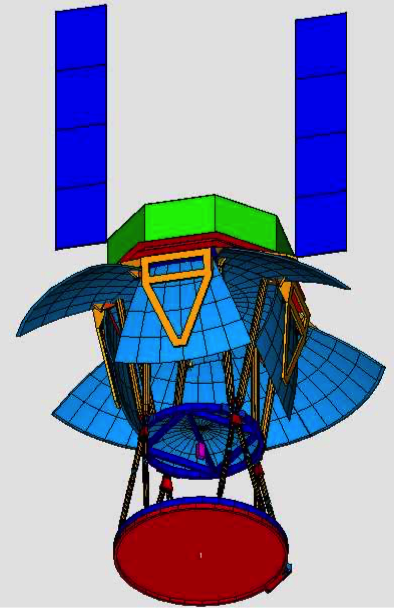}
    \includegraphics[width=0.40\textwidth]{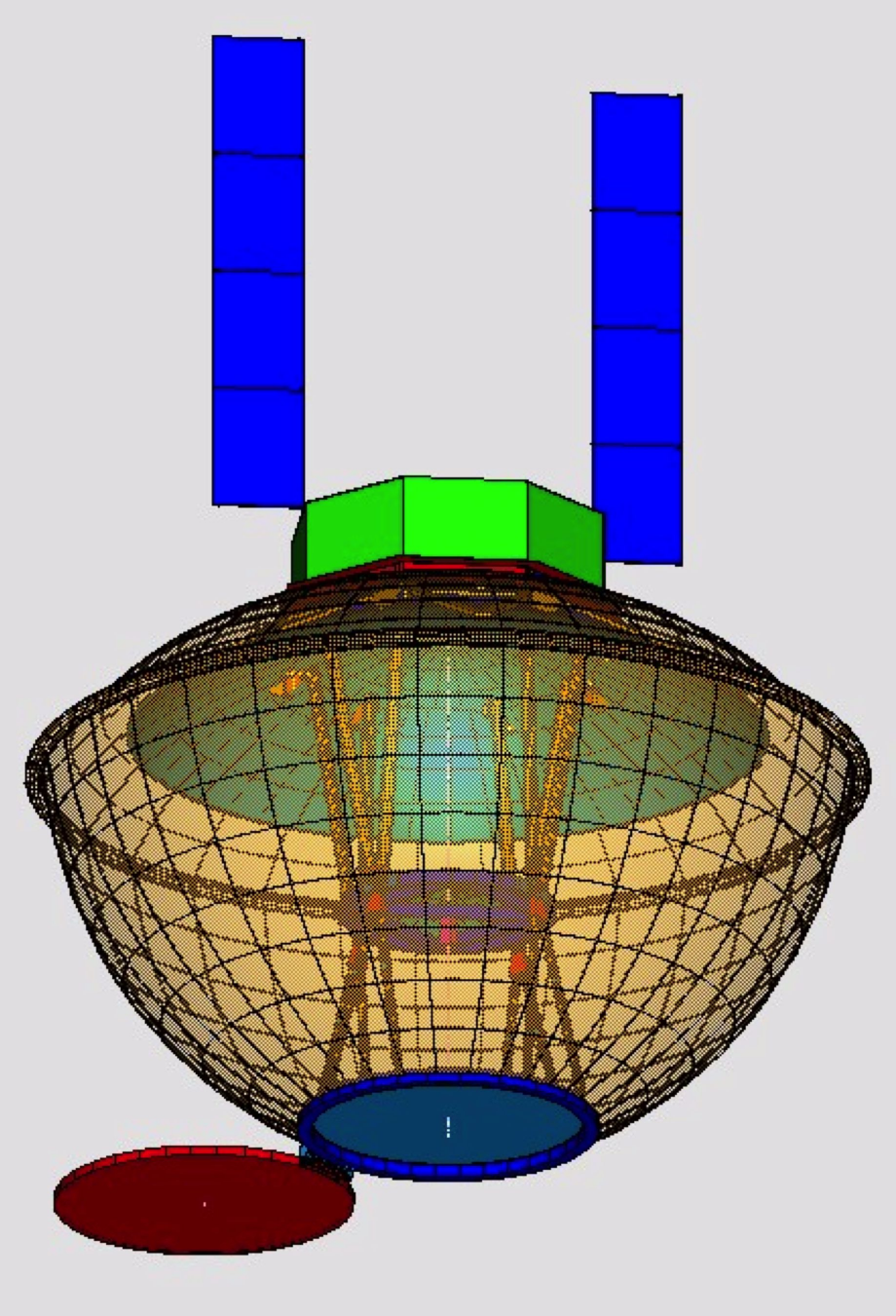}
    \includegraphics[width=0.17\textwidth]{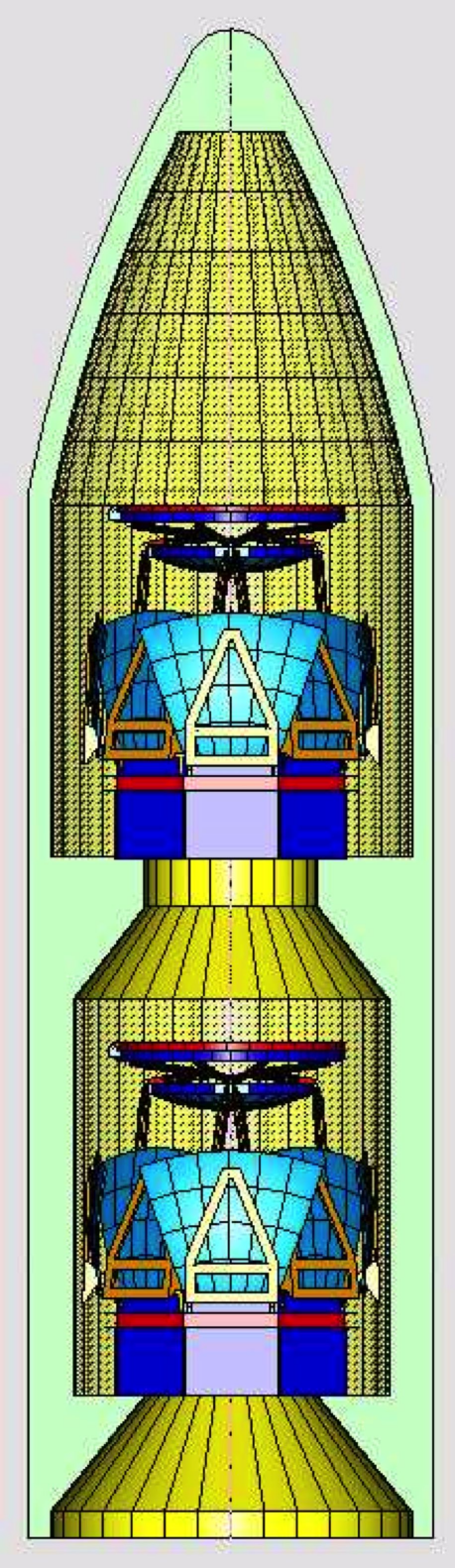}
 \end{center}
\vspace{-0.3cm}
  \caption{Left:  Schematic of an OWL satellite showing the internal, Schmidt telescope structures during deployment, the lightshield is not shown for clarity. Middle: Schematic of an OWL satellite showing the internal, deployed Schmidt telescope, the lightshield is shown as translucent. Right: The stowed OWL spacecraft in a dual-launch fairing on the launch vehicle From Ref. \citenum{OWLidc}.
  }
\vspace{-3 mm}
  \label{OWLsc}
\end{figure}
The charged particles in an EAS produce ionization that excites air fluorescence that is emitted isotropically as the few hundred meter wide (determined by the Moli{\'{e}}re radius, here taken at altitudes below 30 km) and few meter deep 'pancake' of EAS particles moves through the atmosphere close to the speed of light. Thus a 1 km imaging resolution near the ground is more than sufficient to view the vast majority of the EAS in a given snapshot.
A $\mu$-second sampling resolution matches well to that needed to accurately sample the waxing and waning of the EAS longitudinal profile to map the EAS development to determine the UHECR physical properties. That is $\mu$s sampling is close to optimal needed to have a high EAS fluorescence signal to dark-sky airglow background ratio. 
A km image size on the ground from   This imaging requirement of around 1 km from low-Earth orbit, e.g. $400 -- 1000$ km altitude, is very coarse in terms of the that for optical astronomical telescopes\footnote{At 400 nm, 0.1$^\circ$ resolution using a meter-diameter telescope is $\sim 10^4$ away from the diffraction limit.}.  This translates into generous tolerances on the optics, e.g. more similar to a microwave dish than an astronomical telescope.

In the early 2000's, this concept lead to the AIRWATCH concept \cite{1996SPIE.2806..102T} defining a single orbiting telescope for monocular observation of EAS as well as the Orbiting Wide-angle Light collectors (OWL) study, shown conceptually in left panel in Fig.~\ref{OWLstereo}. The two groups worked together in the late 1990's and Fig.~\ref{OWLairW} shows a joint AIRWATCH/OWL logo drawn by John Linsley.  OWL was designed to use the stereo fluorescence EAS technique to more precisely measure the properties of the UHECRs and UHE neutrinos due to the 
superior angular resolution offered by the stereo technique. This provides the ability to achieve  degree angular resolution and $<20\%$ energy resolution, both needed to measure the UHECR nuclear composition evolution. The OWL development included using the air fluorescence technique to search for cosmic neutrinos interacting in the atmosphere and the beamed Cherenkov light from upward-moving EAS from \taon decay source from $\nu_\tau$ interactions in the Earth \cite{Krizmanic2011}.  The OWL instrument and mission were developed at the GSFC Integrated Design Center \cite{IDC} in 2002, and Fig.~\ref{OWLsc} shows schematics of the OWL telescopes and launch configuration based on the IDC work.

%\begin{minipage}

%\begin{figure}{h}
\begin{wrapfigure}{r}{0.51\columnwidth}
\captionsetup{width=0.47\columnwidth}
\vspace{-8mm}
\begin{center}
\includegraphics[width=0.48\textwidth]{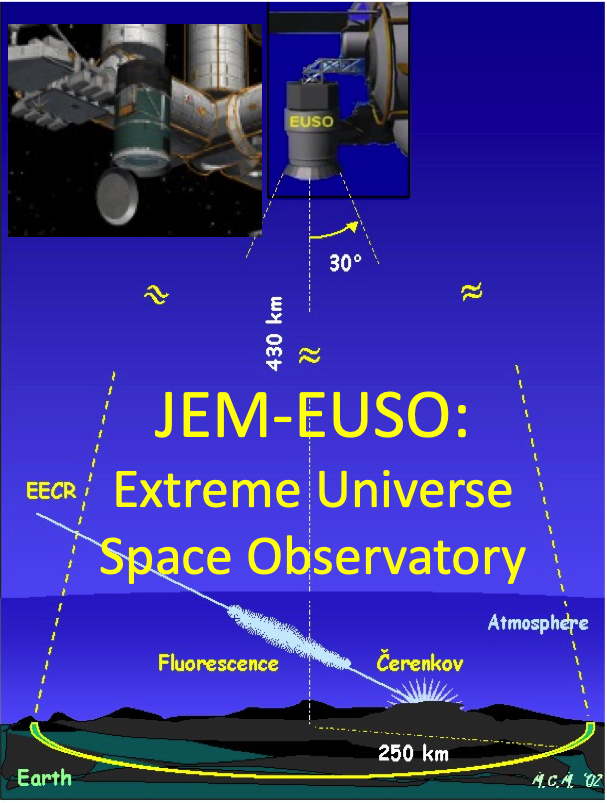}
\end{center}
\caption{Artist sketch of the EUSO on the Japanese Experimental Module (JEM-EUSO) mission proposed as an external International Space Station Payload.  Adapted from Ref. \citenum{2015arXiv150905995V}.
\href{https://arxiv.org/abs/1509.05995}{arXiv:1509.05995}
\label{JEM-EUSO}}
\vspace{-2mm}
\end{wrapfigure}
%\end{figure}

%\end{minipage}

The AirWatch program led to the development of the Extreme Universe Space Observatory (EUSO) program, which included the the EUSO on the Japanese Experiment Module (JEM-EUSO) of the International Space Station (ISS) mission concept \cite{2015ExA....40...19A}. JEM-EUSO was to use
a single nadir-viewing near-UV telescope using refractive, Fresnel optics, to be located on the JEM external facility \cite{2013APh....44...76A}, shown in Fig.~\ref{JEM-EUSO}. The EUSO program included a number of related balloon-borne missions, including EUSO-balloon \cite{2017ICRC...35..445B} with a downward-looking fluorescence telescope using EUSO refractive optics (denoted as EUSO-FT),  the Mini-EUSO experiment using a smaller version of the EUSO-FT \cite{2021ApJS..253...36B} on the ISS, EUSO-SBP \cite{2019NIMPA.936..237O} ULDB payload flown in 2017 with a EUSO-FT, and the EUSO-SPB2 ULDB mission \cite{2020NIMPA.95862164S} planned to fly in the spring of 2023, that includes a downward-looking Schmidt fluorescence telescope and an Earth-limb viewing Schmidt Cherenkov telescope to search for upward tau neutrino induced EAS and observe Cherenkov light from VHECRs viewed over-the-limb. 

\section{Space-based UHECR Missions}

\begin{figure}
\vspace{-3mm}
 \begin{center}
    \includegraphics[width=0.99\textwidth]{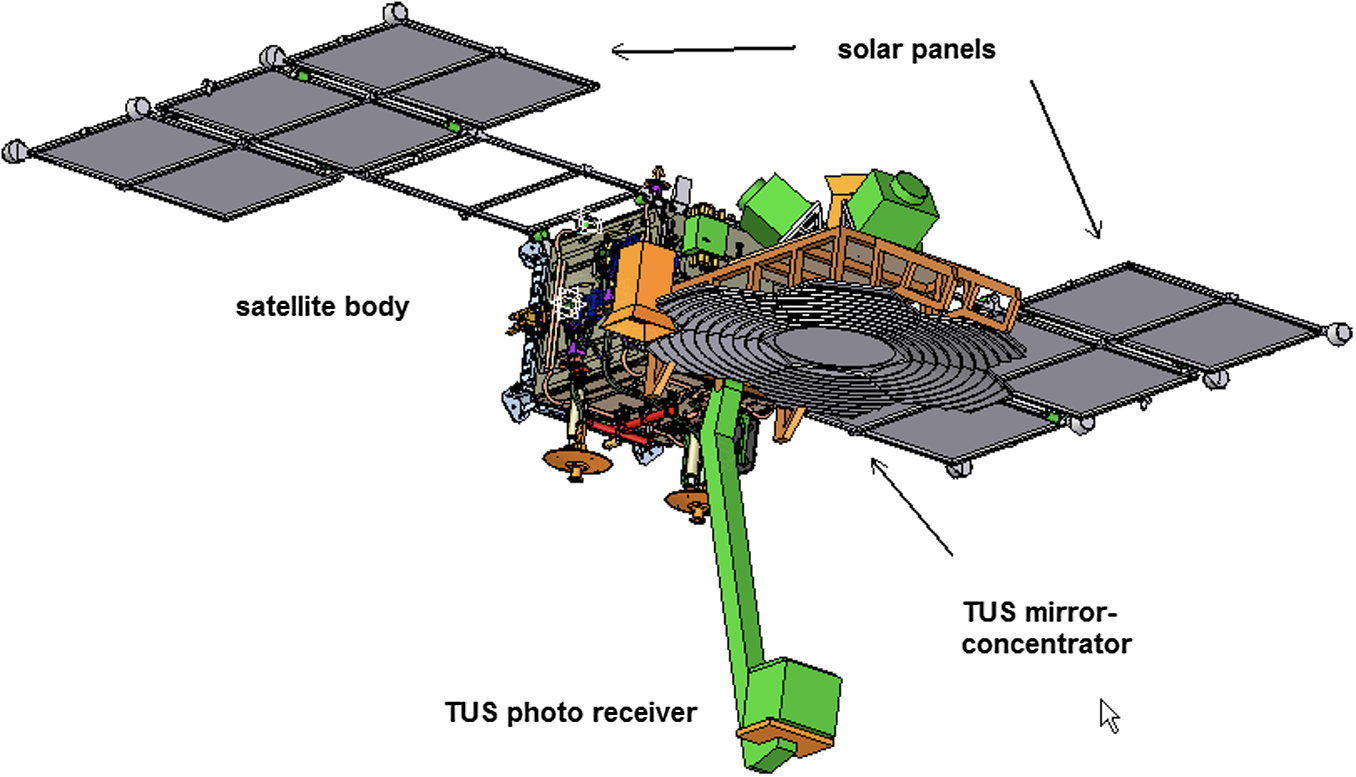}
  \end{center}
  \vspace{-1mm}
  \caption{The schematic of the TUS experiment on the Lomonosov satellite. The Fresnel mirror, with $\sim$2 m$^2$ area, has a field of view of 9$^\circ$ imaging EAS in a focal plane with 256 pixels. Three years of operation at $\sim 500$ km altitude lead to an exposure of 12,000 km$^2$ sr yr for $E_{\rm UHECR} > 300$ EeV. From Ref.~\citenum{2017SSRv..212.1687K}.
  DOI: \href{https://doi.org/10.1007/s11214-017-0403-3}{10.1007/s11214-017-0403-3}
  \label{TUS}}
  \vspace{-1mm}
%\end{wrapfigure}
\end{figure}

\begin{figure}[ht]
\begin{center}
    \includegraphics[width=0.35\textwidth]{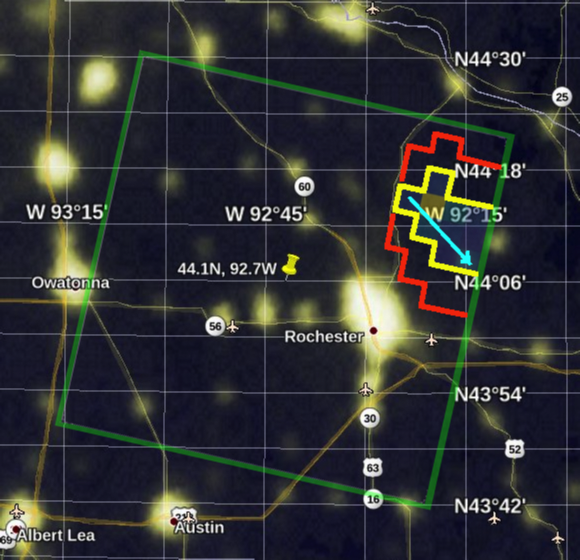}
    \includegraphics[width=0.55\textwidth]{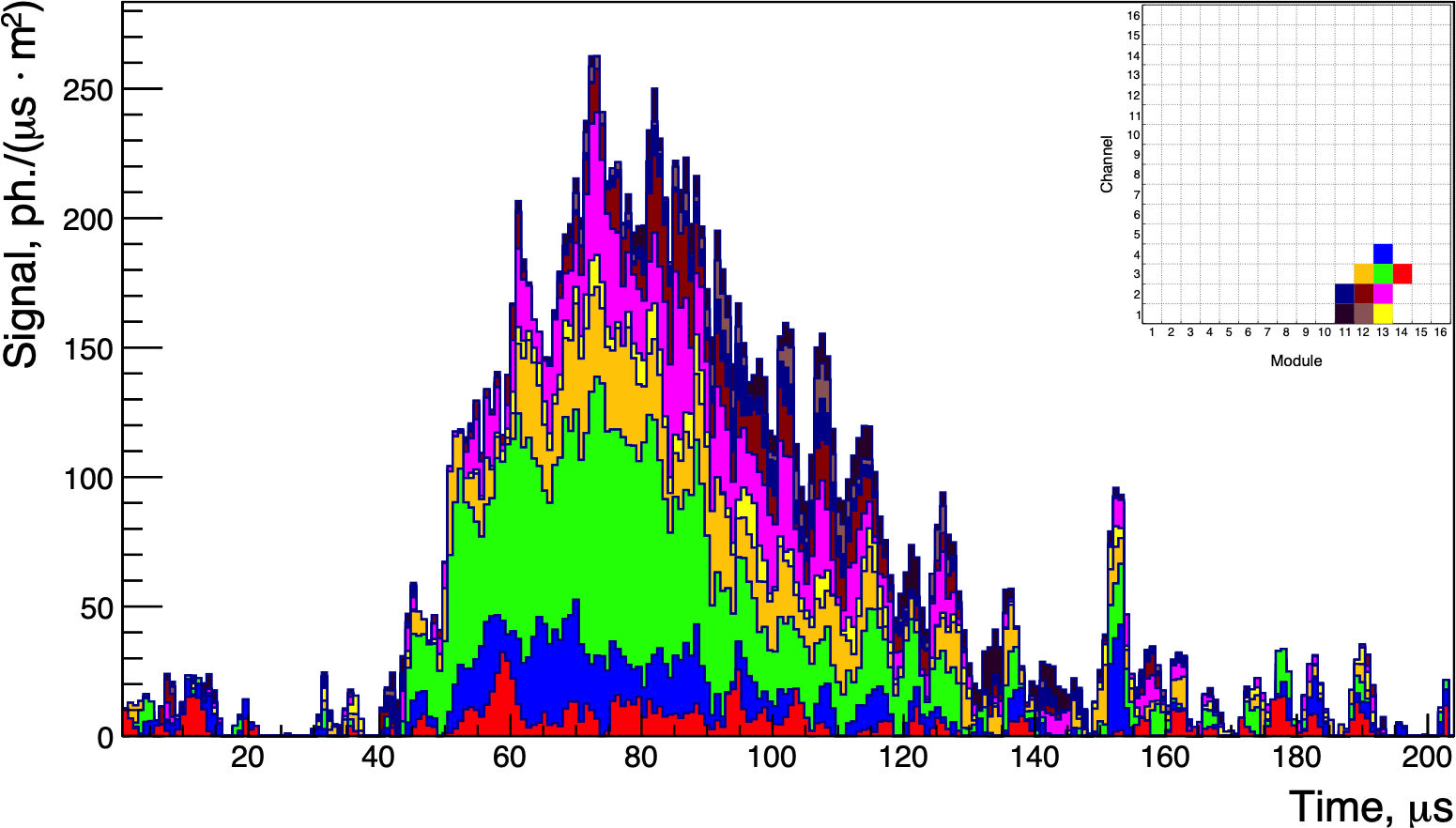}
  \end{center}
\vspace{-0.3cm}
  \caption{Left: The FoV of TUS (green square) for the TUS161003 event overlaid on a Google Earth\footnote{https://google.com/earth/} map that includes NASA night Earth data. The yellow area shows the focal plane hit map while the red areas include the area with the effects of timing errors. Right: The light curve stacking the 10 hit channels for the TUS161003 event. The inset shows the location of the 10 hit pixels. From Ref.~\citenum{2020JCAP...03..033K}.
  DOI: \href{https://doi.org/10.1088/1475-7516/2020/03/033}{10.1088/1475-7516/2020/03/033}}
\vspace{-3 mm}
  \label{TUSevent}
\end{figure}

Somewhat independently, the Tracking Ultraviolet Set-up (TUS) experiment \cite{2017SSRv..212.1687K, Klimov:2017dul}, shown in Fig.~\ref{TUS} was built, launched, and operated on the Lomonosov satellite from May 19, 2016 until May 30, 2019. Three years of operation at $\sim 500$ km altitude in a sun-synchronous orbit lead to an exposure of 12,000 km$^2$ sr yr for $E_{\rm UHECR} > 300$ EeV \cite{2017SSRv..212.1687K}. TUS employed an optical system with a $\sim$2 m$^2$ Fresnel mirror, a field-of-view of 9$^\circ$, and imaged the near-UV air fluorescence emission from EAS in a focal plane with 256 pixels.  The design did not include a light shroud and this lead to more sensitivity to optical backgrounds. The relatively modest size of the optical light collection led to a UHECR detection energy threshold of $\gsim 70$ EeV, but TUS did detect six UHECR EAS candidate events whose origin may be due to background \cite{2020JCAP...03..033K, 2021ICRC...37..316K}. Fig.~\ref{TUSevent} shows the location and light curve for an UHECR candidate event observed over Minnesota in 2016. While the UHECR candidate sample is small, TUS has measured a wealth of low-intensity transient atmospheric phenomena \cite{2019RemS...11.2449K}, including transient luminous events (TLEs), meteors, and structures in the aurora \cite{2021JASTP.22005672K}, the last taking advantage of the Lomonosov satellite near-polar orbit. 

\begin{figure}[ht]
\begin{center}
  \includegraphics[width=0.54\textwidth]{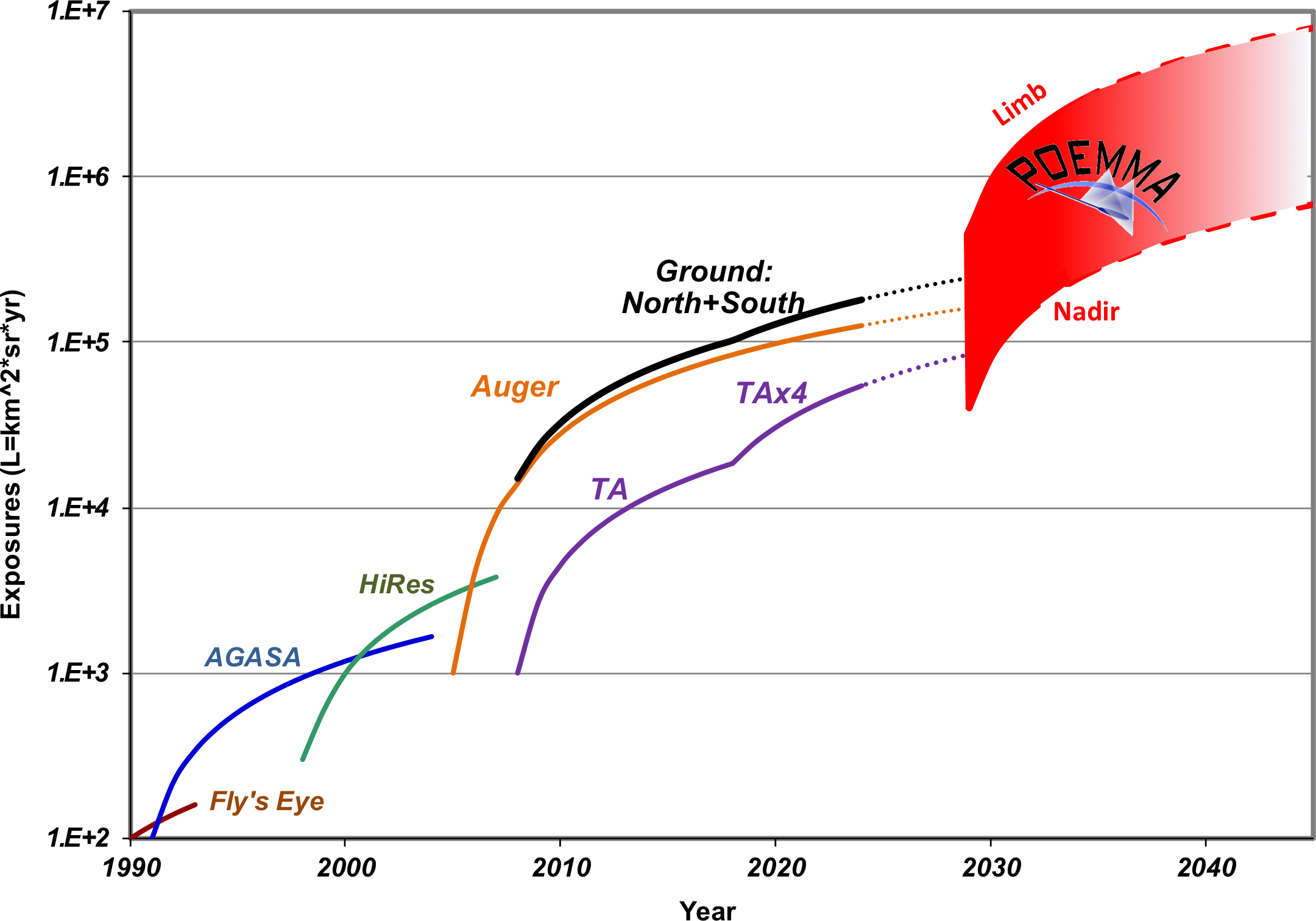}
  \hspace*{2 pt}
   \includegraphics[width=0.42\textwidth]{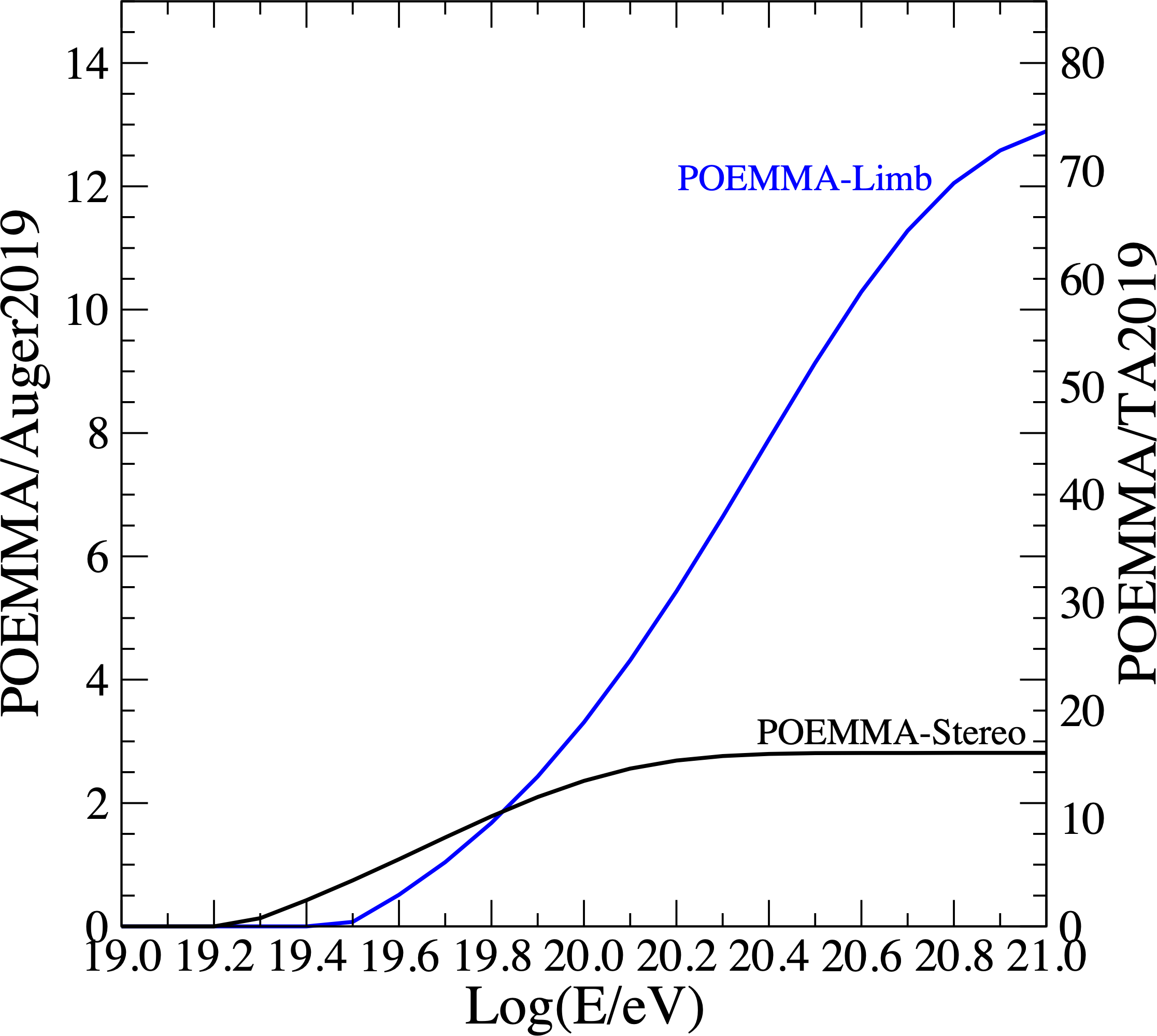}
\end{center}
    \caption{Left: The anticipated UHECR exposure growth curve for POEMMA compared to other UHECR experiments assuming nadir-pointing stereo fluorescence measurements versus limb-pointing UHECR measurements, but with less precision. Right: The comparison of 5-year POEMM exposure versus UHECR energy in terms of the PAO and TA exposures as of the 2019 ICRC. 
From Ref. \citenum{2021JCAP...06..007P}.
  DOI:\href{https://doi.org/10.1088/1475-7516/2021/06/007}{10.1088/1475-7516/2021/06/007}}
\label{POEMMAexposure}
\end{figure}

The OWL study along with the ample work of the EUSO program, especially the MAPMT-based hardware for air fluorescence telescopes, led to the development in 2018  of the Probe of MultiMessenger Astrophysics (POEMMA), designed as a NASA Astrophysics Probe-class mission.  POEMMA was selected as one of several different astrophysics missions to be developed under study program (NASA NNH16ZDA001N-APROBES) to help define the NASA Astrophysics Probe class, which is larger than an Explorer class but smaller than a Flagship mission. More importantly it provided resources to develop the POEMMA instruments, spacecraft, and mission designs in both the Instrument Design Lab (IDL) and Mission Design Lab (MDL) at the IDC \cite{IDC} at NASA/GSFC. Thus the POEMMA study provided the opportunity for the UHECR community to develop as optimal a UHECR mission as possible, considering the relatively generous constraints defined for the probe class mission. The POEMMA UHECR exposure performance is shown in Fig.~\ref{POEMMAexposure}. 

\begin{figure}[h]
%\addtocounter{figure}{1}
\begin{center}
    \includegraphics[width=0.69\textwidth]{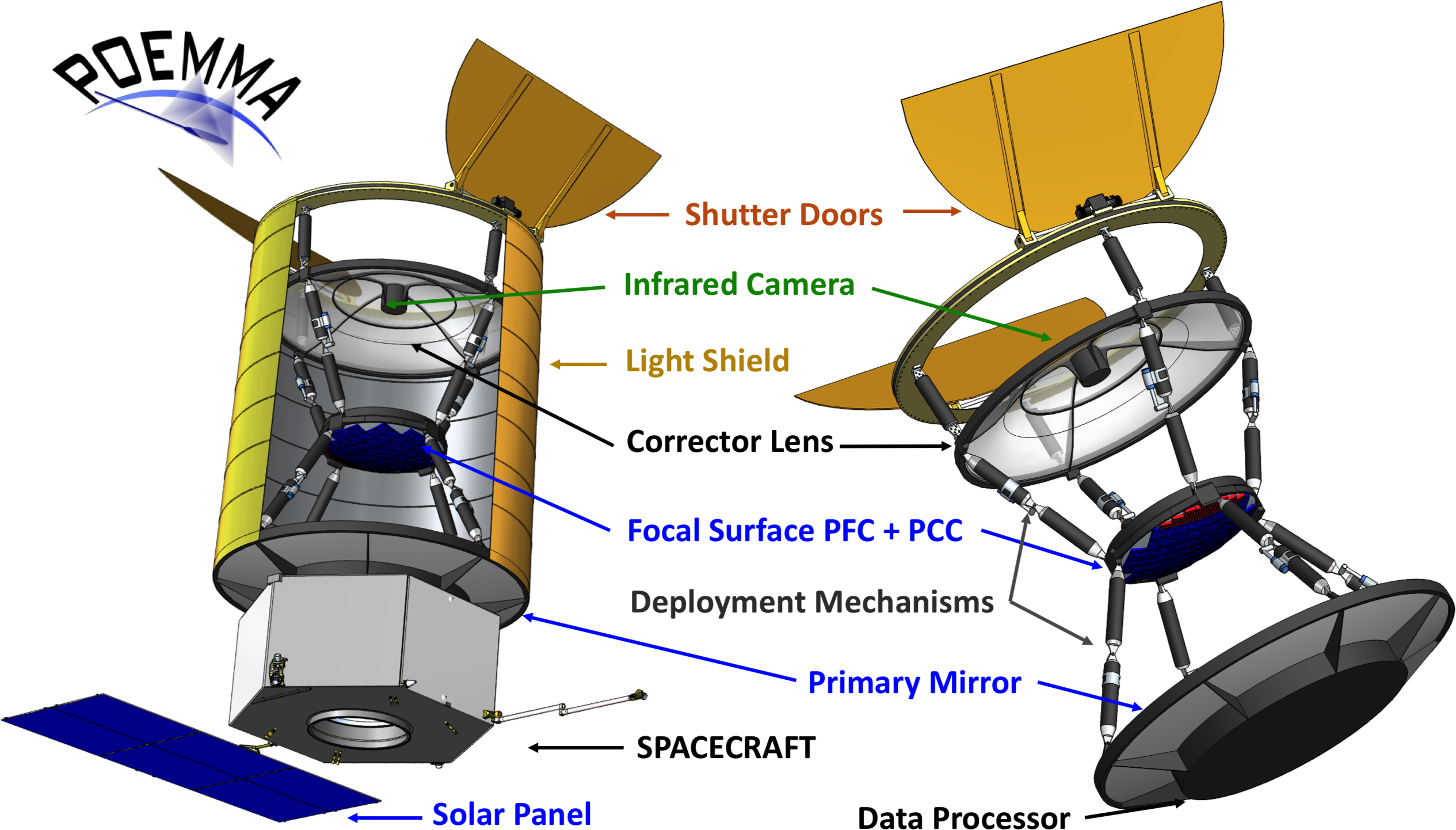}
    \hspace*{2 pt}
        \includegraphics[width=0.24\textwidth]{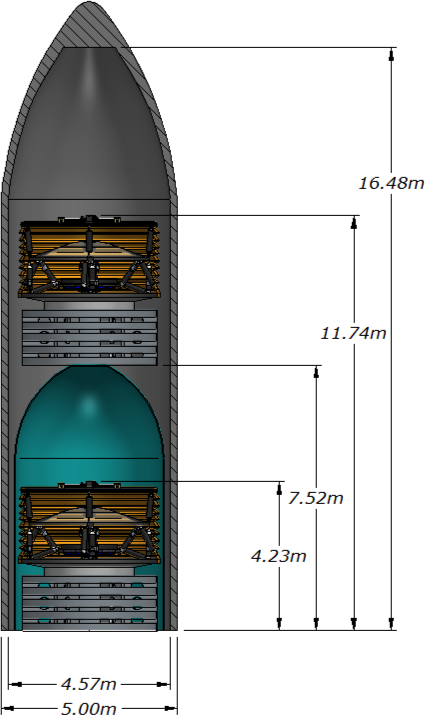}
 \end{center}
\vspace{-0.3cm}
  \caption{Schematics of a POEMMA satellite (left) and the Schmidt telescope (right) consisting of a 4-m diameter primary mirror, 3.3-m diameter corrector plate, and 1.6-m diameter focal surface comprised of 126,720 pixels in the POEMMA Fluorescence Camera (PFC) and 15,360 pixels in the POEMMA Cherenkov Camera (PCC)  Several components are detailed in the schematic including infrared cameras which will measure cloud cover within the 45$\circ$ full FoV of each telescope during science observations. Right: Right: The stowed POEMMA spacecraft in a dual-launch fairing on the launch vehicle with length dimensions. From Ref. \citenum{2021JCAP...06..007P}.
  DOI:\href{https://doi.org/10.1088/1475-7516/2021/06/007}{10.1088/1475-7516/2021/06/007}
 }
%\addtocounter{figure}{-1}
%\vspace{-3 mm}
  \label{POEMMAinst}
\end{figure}

\begin{figure}[h]
\begin{center}
    \includegraphics[width=0.99\textwidth]{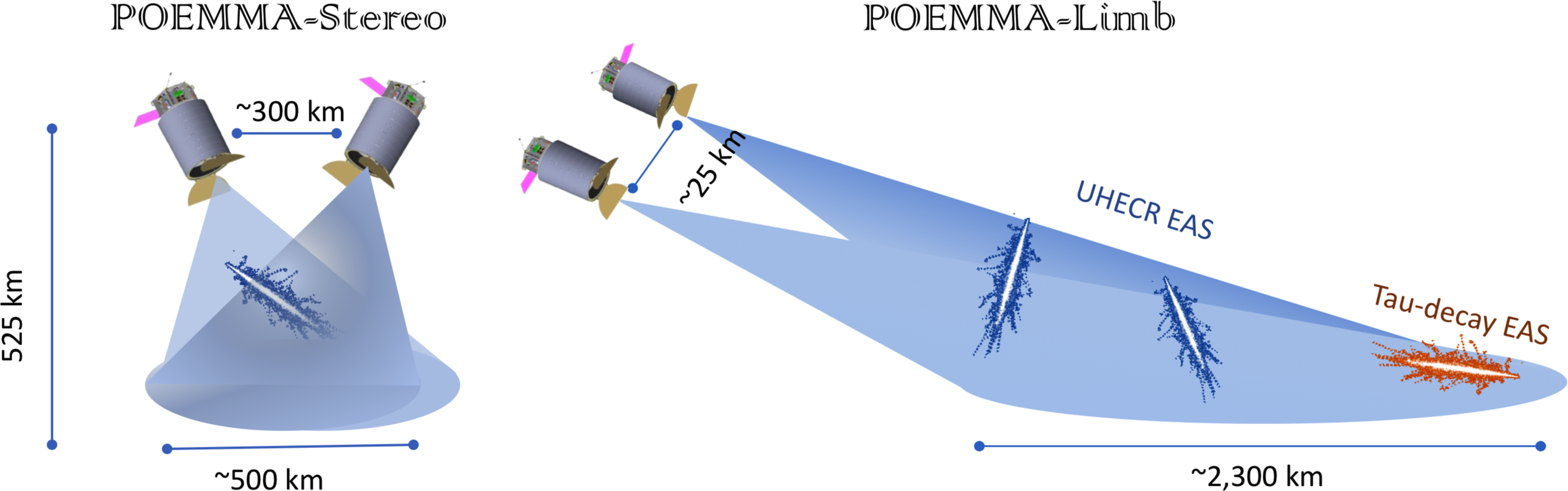}
 \end{center}
\vspace{-0.3cm}
  \caption{The POEMMA science modes. Left: POEMMA-Stereo where the spacecraft are separated and viewing a common atmospheric volume to measure the fluorescence emission from EAS. Right: POEMMA-Limb where the instruments are tilted to view near and below the limb of the Earth for optical Cherenkov EAS induced by tau neutrino events in the Earth. From Ref. \citenum{2021JCAP...06..007P}. 
  DOI:\href{https://doi.org/10.1088/1475-7516/2021/06/007}{10.1088/1475-7516/2021/06/007}
  }
%\vspace{-3 mm}
  \label{POEMMAmodes}
\end{figure}

POEMMA is designed to observe cosmic neutrinos and to identify the sources of UHECRs with full-sky coverage. POEMMA consists of two spacecraft that co-view EAS while flying in a loose formation, separated by 300 km, at 525 km altitudes at 28.5$^\circ$ inclination. Each spacecraft hosts a large Schmidt telescope with a novel focal plane optimized to observe both the isotropic near-UV fluorescence signal generated by EAS from UHECRs and UHE neutrinos and forward beamed, optical Cherenkov signals from EAS. Fig.~\ref{POEMMAinst} shows the schematics of each spacecraft and telescope and launch vehicle integration developed during the POEMMA probe study \cite{POEMMAnasaStudy}. In neutrino limb-viewing Cherenkov mode, POEMMA will be sensitive to cosmic tau neutrinos above 20 PeV by observing the upward-moving EAS induced from tau neutrino interactions in the Earth. The POEMMA spacecraft are designed to quickly re-orient to a Target-of-Opportunity (ToO) neutrino mode to view and follow transient astrophysical sources with unique flux sensitivity over the full sky.
Each POEMMA Schmidt telescope is comprised of a 4-meter diameter primary mirror and 3.3-meter diameter corrector lens that yields nearly 6 m$^2$ effective on-axis light collecting ability to both perform precision UHECR measurements and search for UHE cosmic neutrinos. The large collection area provides for a UHECR detection threshold energy of 20 EeV, which is needed to have sufficient overlap with ground-based measurements performed by the Pierre Auger Observatory (PAO: in the southern hemisphere) \cite{ABRAHAM201029,ABRAHAM2010227} and the Telescope Array (TA: in the northern hemisphere) \cite{ABUZAYYAD201287,TOKUNO201254}.  Furthermore, UHECR measurements are required for $E_{\rm CR} \approx 40$ EeV to be able to measure the anisotropic features measured by PAO (correlation to star-burst galaxies) \cite{Abreu:2021eL} and TA (hot-spot) \cite{Kim:2021Aj}. That is an experiment that has high UHECR sensitivity will also have high sensitivity for UHE neutrino interactions that occur deep in the atmosphere that are well separated from UHECR EAS. \cite{2021JCAP...06..007P}. The high-statistics ($\ge 1,400$ events for a five-year mission) full-sky UHECR measurements above 20 EeV using the stereo air fluorescence technique would provide a major advance in discovering the sources of UHECRs \cite{2021JCAP...06..007P,2010PhRvD..81l3009A}. POEMMA also provides unique sensitivity to UHE cosmic neutrino searches using stereo air fluorescence measurements, and an Earth limb-pointed mode to observe VHE Earth-interacting cosmic tau neutrinos using the beamed optical Cherenkov light generation from EAS for $E_\nu \gsim$ 20 PeV \cite{2019PhRvD..99f3011R,2020PhRvD.102l3013V}. Fig.~\ref{POEMMAmodes} illustrates the two science modes of POEMMA while Tab.~\ref{UHECRtab} summarizes the UHECR performance.


\begin{table}
\centering
\caption{Overview of the simulated POEMMA UHECR measurement capabilities \cite{2020PhRvD.101b3012A}. Adapted from Ref \cite{2021NIMPA.98564614K}.}
\label{UHECRtab}       % Give a unique label
% For LaTeX tables you can use
\begin{tabular}{ll}
\hline
% Optics & Schmidt Telescopes &  \\\hline
Parameter &  Performance   \\ \hline
 UHECR Stereo & 260,000 km$^2$ sr (50 EeV) \\
 Geometry Factor & 400,000 km$^2$ sr ($\ge 100$ EeV) \\ \hline
 Obs Duty Cycle & 13\%  \\ \hline
 Physics Energy Thres & 20 EeV \\ \hline
UHE Stereo Energy Res & $<19\%$ (50 EeV) \\ \hline
UHE Angular Res & $\le 1.2^\circ$ (50 EeV) \\ \hline
UHE \xmax~ Res & $\le 30$ g/cm$^2$ (50 EeV) \\ \hline
CR rejection factor
%& \\
for UHE neutrinos & $2 \times 10^{-4}$ \\ \hline
Sky Coverage & $\pm 20\%$ @ 50 EeV (1 Year) \\
Variability &  $\pm 10\%$ @ 50 EeV (5 Year) \\ \hline
 \end{tabular}
% Or use
%\vspace*{5cm}  % with the correct table height
\end{table}

\begin{table}
%{\small
\centering
\caption{POEMMA Observatory Specifications: Observatory = Two Telescopes; Each Telescope = Instrument + Spacecraft. Adapted from Ref. \citenum{2021JCAP...06..007P}.
    DOI:\href{https://doi.org/10.1088/1475-7516/2021/06/007}{10.1088/1475-7516/2021/06/007}}
\label{tab-1}  \
\begin{tabular}{lll}
\hline
\hline
Telescope: & Instrument &   \\ \hline
Optics &  Schmidt & 45$^\circ$ full FoV  \\
 & Primary Mirror & 4 m diam. \\
 & Corrector Lens & 3.3 m diam.  \\
 & Focal Surface & 1.6 m diam.  \\  
 & Pixel Size & $3 \times 3$ mm$^2$   \\ 
& Pixel FoV & 0.084$^\circ$  \\
PFC & MAPMT (1$\mu$s)& 126,720 pixels   \\
PCC & SiPM (20 ns)& 15,360 pixels   \\ \hline
Observatory & \multicolumn{2}{l}{Each Telescope}  \\ \hline
 & Mass & 1,550 kg  \\
 & Power (w/cont) & 700 W    \\
 & Data & $<$ 1 GB/day  \\
\hline
  & Spacecraft  & \\ \hline
 & Slew rate & 90$^\circ$  in 8 min \\
 & Pointing Res. & 0.1$^\circ$ \\
 & Pointing Know. & 0.01$^\circ$ \\
 & Clock synch. & 10 ns \\  
   & Data Storage & 7 days \\ 
& Communication & S-band \\
 & Wet Mass & 3,450 kg \\
& Power (w/cont)& 550 W \\ \hline
  &Mission  & (2 Telescopes)\\ \hline
  & Lifetime & 3 year  (5 year goal)\\
 & Orbit & 525 km, 28.5$^\circ$ Inc \\
 & Orbit Period & 95 min \\
 & Telescope Sep. & $\sim$25 - 1000 km
\\\hline \hline
\end{tabular}
\label{POEMMAspecs}
\end{table}

%The charged particles in an EAS produce ionization that excites air fluorescence that is emitted isotropically  as the few hundred meter wide (determined by the Moli{\'{e}}re radius, here taken at altitudes below 10 km) and few meter deep 'pancake' of EAS particles moves through the atmosphere close to the speed of light.
While the wavelength band for air fluorescence extends from below 200 nm to over 1000 nm \cite{1964JChPh..41.3946D}, the vast majority is in the band from 300 nm to 450 nm \cite{1964JChPh..41.3946D,1967PhDT........28B}. This allows for the use of UV filters to limit the effects of the dark-sky airglow background to be  $\sim$ 500 photons m$^{-2}$ ns$^{-1}$ $sr^{-1}$ \cite{2005APh....22..439B}, and thus help reduce the UHECR EAS observational energy threshold. 
POEMMA in 525 km altitude low Earth orbit (LEO), will view the air fluorescence EAS signals using a large $45^\circ$ full field-of-view (FoV). Thus a vast amount of atmosphere will be  monitored, i.e. $\sim 10^{13}$ metric tons \cite{OWL,2021JCAP...06..007P}. The EAS imaging requirements that correspond to resolving approximately 1 km spatial lengths on the Earth's surface leads to a  iFoV$=0.084^\circ$ for each pixel in the POEMMA focal plane. This iFoV from 525 km altitude LEO, sets the integration time of the imaged air fluorescence to be 1 $\mu$s to optimize the signal to dark-sky background ratio for EAS fluorescence detection.

\begin{wrapfigure}{r}{0.55\columnwidth}
%\addtocounter{figure}{-1}
\captionsetup{width=0.5\columnwidth}
%\begin{figure}[h]
\centering
\includegraphics[width=0.5\columnwidth]{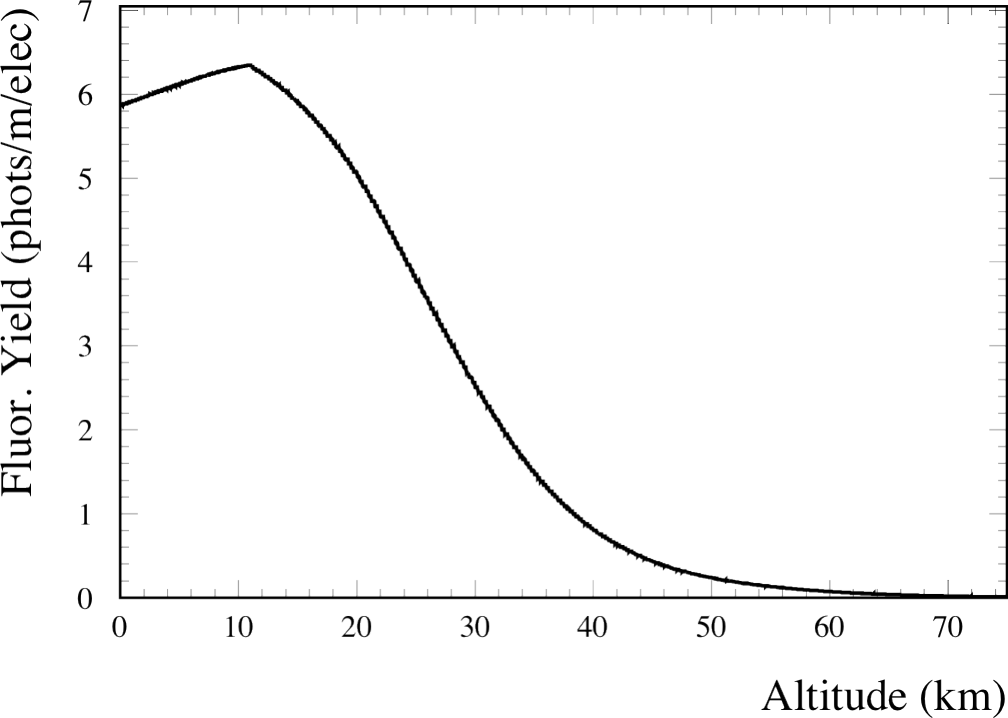}
\caption{The air fluorescence yield as a function of altitude. From Ref.~\citenum{2021NIMPA.98564614K}.
  DOI: \href{https://doi.org/10.1016/j.nima.2020.164614}{10.1016/j.nima.2020.164614}
  }
\label{FYalt}       % Give a unique label
%\addtocounter{figure}{1}
%\end{figure}
\end{wrapfigure}

The yield of fluorescence light (at STP) is of order of 10 photons per MeV of ionization energy deposited by EAS particles \cite{PDG}. 
For EAS viewed from space using a wide FoV telescope, the entire EAS development is within the FoV.  Fig.~\ref{FYalt} shows the air fluorescence yield as a function of altitude \cite{2021NIMPA.98564614K}, and shows that the majority of the air fluorescence signal is generated below an altitude of $\lsim $30 km.
Viewed from a distance of more than 10 km, the EAS can be well approximated as a line along which the shower progresses, at first brightening then diminishing. A parametric model of the longitudinal development using the Gaisser-Hillas function \cite{Gaisser-Hillas-1977-ICRC-15-8-353} along the shower track combined with ray-tracing of the generated isotropic light provides input for Monte Carlo simulations for the generation of EAS, optical (and radio) signals, propagation and attenuation through the atmosphere, and modeling the detection of a space-based instrument \cite{1999ICRC....2..388K,1999ICRC....1..445M,2010APh....33..221B}. The leftmost schematic in Fig.~\ref{POEMMAmodes} illustrates stereo fluorescence detection using two Schmidt telescopes for POEMMA \cite{2021JCAP...06..007P}.

The EAS also creates an Cherenkov light signal due to the fact the charged particles are moving faster than the speed of light in the medium, e.g. the atmosphere. This Cherenkov light can scatter in the atmosphere and also be detected, but tends to be a small fraction of the total EAS optical signal, $\sim$ 10\% based on space-based UHECR simulation studies.  The beamed Cherenkov light can also scatter off the ground and be used to determine the ground-spot of the UHECR trajectory (see Fig.~\ref{JEM-EUSO}, which improves the accuracy in the reconstruction of the EAS trajectory \cite{2010APh....33..221B}.

\subsection{Optical Detection Instrumentation}

\begin{figure}[t]
\begin{center}
    \includegraphics[width=0.29\textwidth]{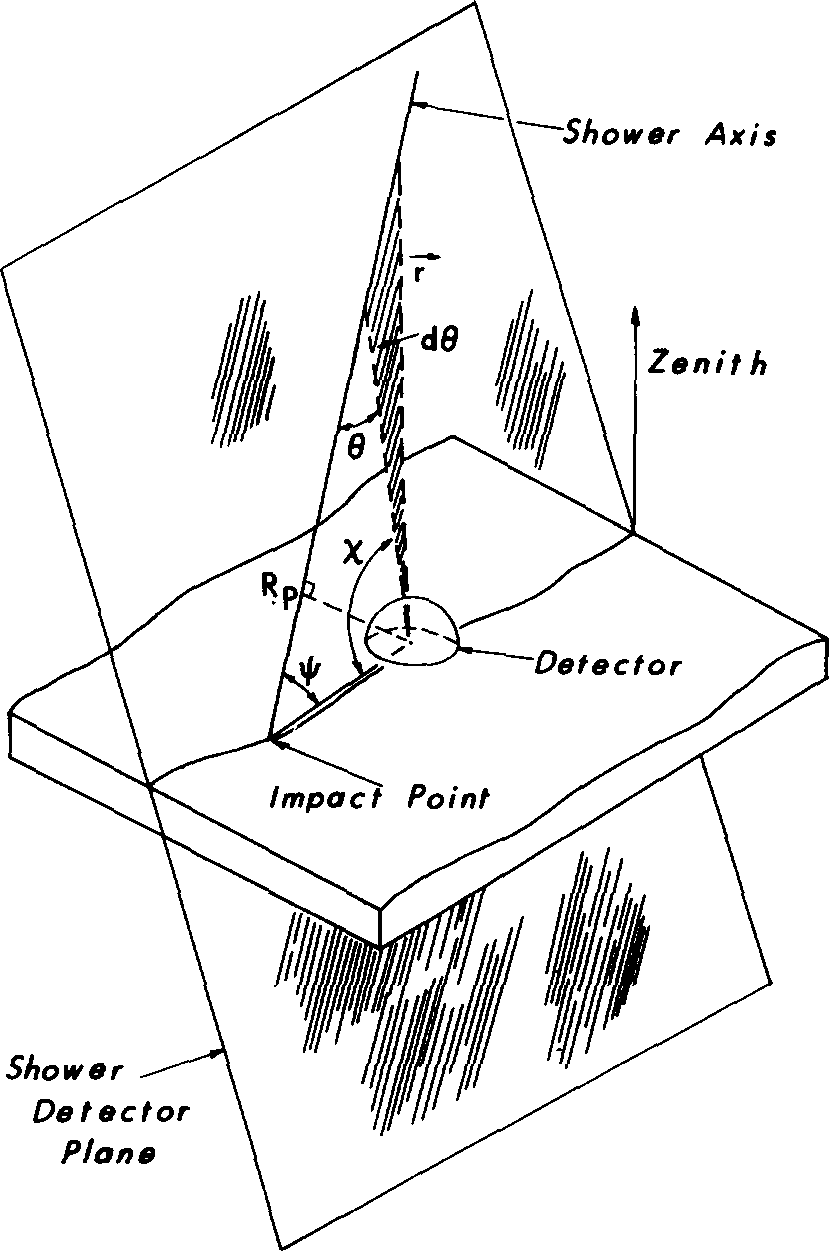}
    \hspace*{1 pt}
   \includegraphics[width=0.65\textwidth]{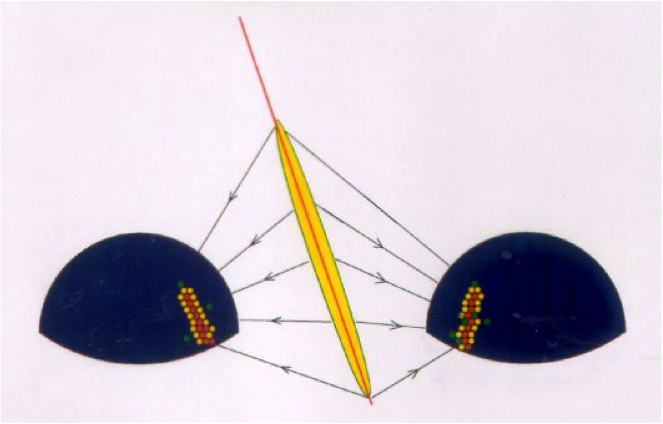}
 \end{center}
\vspace{-0.3cm}
  \caption{Left: The UHECR reconstruction using a single site measurement of the fluorescence from and EAS. Right: Stereo observations of an EAS illustrating how the two EAS-detector planes intersect to form a line that defines the EAS trajectory. Left: From Ref. \citenum{1985NIMPA.240..410B}.       DOI:\href{https://doi.org/10.1016/0168-9002(85)90658-8}{10.1016/0168-9002(85)90658-8}
  Right: From Ref. \citenum{UtahWebSite}.
}
\vspace{-3 mm}
  \label{FlysEye}
\end{figure}
\begin{figure}[t]
\begin{center}
    \includegraphics[width=0.99\textwidth]{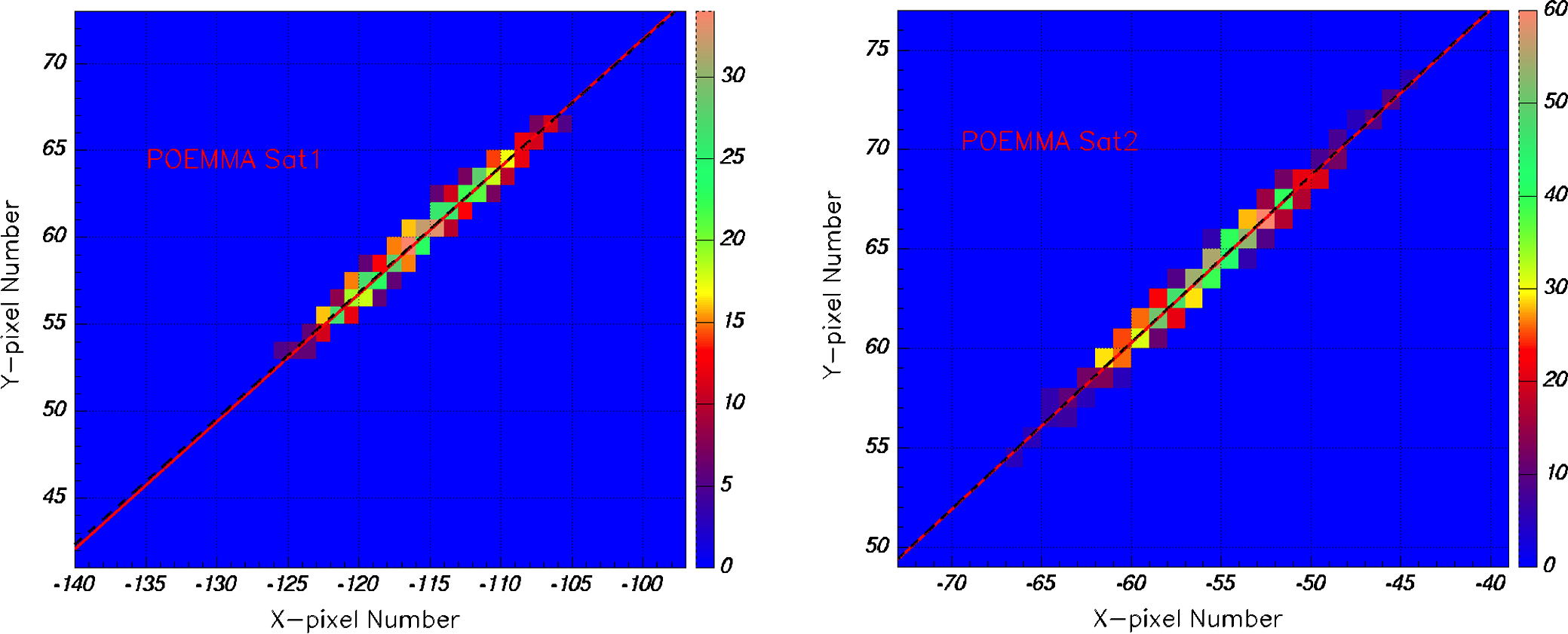}
 \end{center}
\vspace{-0.3cm}
  \caption{A co-viewed, stereo reconstructed 50 EeV simulated UHECR in the two POEMMA focal planes. The solid line shows the simulated trajectory while the dashed line is defined by the reconstructed trajectory. The color map provided the photo-electron statistics simulated for each pixel. From Ref. \citenum{2020PhRvD.101b3012A}.
      DOI:\href{https://doi.org/10.1103/PhysRevD.101.023012}{10.1103/PhysRevD.101.023012}
}
\vspace{-3 mm}
  \label{EASstereo}
\end{figure}

The stereo fluorescence technique, pioneered by the ground-based Fly's Eye experiment in Utah\cite{1985NIMPA.240..410B} provides a experimental methodology to precisely measure the development of EAS thus achieving good angular, energy, and \xmax~ resolutions for each EAS measurement. Fig.~\ref{FlysEye} illustrates two different EAS reconstruction techniques. Monocular reconstruction uses a single detector location to measure the temporal and spatial evolution of the EAS in a segmented UV camera, using the timing and geometry to determine the EAS trajectory and location of \xmax. The second technique uses two (or more) EAS observations and stereo reconstruction to provide a more precise determination of the 3-dimensional location of the EAS development and thus more precise angular, energy, and \xmax~ measurements.  When applied from a space-based experiment, these techniques allow for an immense amount of atmosphere to be used as a UHECR and UHE neutrino detector. In particular, the stereo reconstruction technique using an iFoV $\approx 0.1^\circ$ from a LEO altitude of 525 km, precisely defines the EAS 3-dimensional trajectory. Fig.~\ref{EASstereo} shows the simulation results for a 50 EeV UHECR proton event as seen in the focal plane of the POEMMA telescopes, with the reconstructed and simulation MC-truth tracks superimposed on the pixel heat maps.  The fact that these are virtually indistinguishable demonstrate the accuracy provided by using two well-separated EAS detector planes. This provides a straightforward geometrical method to determine the line where the two planes overlap, thus accurately determining the location of the EAS. Only in the case where the angle between the planes is small, e.g. $\lsim 5^\circ$, does the pure geometrical reconstruction has difficulty \cite{2020PhRvD.101b3012A}.  Timing information can be used to reconstruct this call of UHECR events.
This leads to exception angular resolution ($\sigma \lsim 1.5^\circ$ above 20 EeV) of the EAS which is needed to obtain the required energy and \xmax~ resolutions to accurately measure the UHECR spectrum and UHECR nuclear composition evolution  \cite{2020PhRvD.101b3012A}. Furthermore, the ability to precisely reconstruct the EAS trajectory, energy, and \xmax~ provides a method to separate the more horizontal neutrino events from the much more copious UHECR events.
For POEMMA, the selection of $X_{\rm Start} \gsim 2000 ~{\rm g/cm}^2$ has been shown, based on the analysis of the $X_{\rm Start}$ distributions for UHECR protons for $E_{\rm CR} \ge 40$ EeV that leads to a POEMMA response with a $\gsim 10^4$ rejection factor for UHECR background events\cite{2020PhRvD.101b3012A}.  These UHE neutrino measurements are performed whenever the POEMMA satellites are in stereo-fluorescence UHECR operational mode.
It is noted that Earth-emergent EAS from neutrino interactions in the Earth also provide a fluorescence signal that can be observed from space if the instruments are pointed to the Earth's limb  \cite{2015arXiv150905995V}. 

%The most advanced current design for space-based UHECR and VHE/UHR cosmic neutrino detection experiment is Probe of MultiMessenger Astrophysics (POEMMA) \cite{2021JCAP...06..007P}. A NASA Astrophysics Probe-class mission,  

\begin{figure}[h]
\begin{center}
    \includegraphics[width=0.99\textwidth]{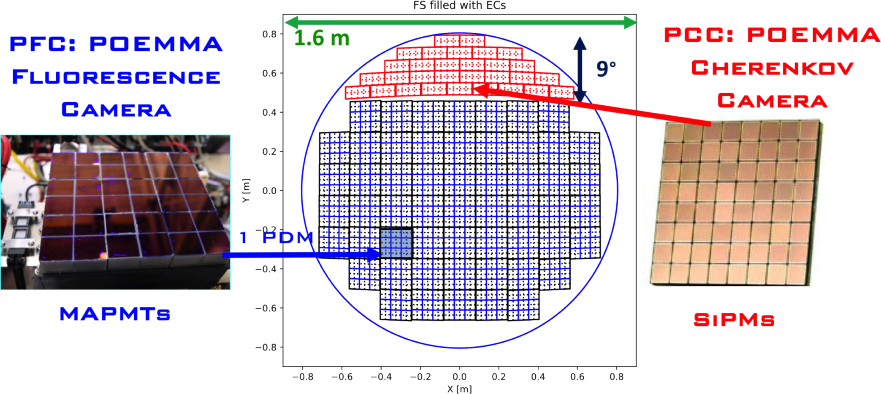}
 \end{center}
\vspace{-0.3cm}
  \caption{The layout of the hybrid focal plane of a POEMMA Schmidt telescope. The majority of the area is comprised of PFC MAPMT modules with a UV filter to record the 300$-$500 nm air fluorescence light in 1 $\mu$s snapshots. The PCC is comprised of SiPM pixels whose 300$-$1000 nm wavelength response is well-matched to that from the EAS optical Cherenkov signals and are recorded with 10 ns cadence. From Ref. \citenum{2021JCAP...06..007P}.
    DOI:\href{https://doi.org/10.1088/1475-7516/2021/06/007}{10.1088/1475-7516/2021/06/007}
  }
\vspace{-3 mm}
  \label{POEMMAfp}
\end{figure}

The specifications of the POEMMA Observatory, spacecraft, and telescopes are detailed in Tab.~\ref{POEMMAspecs}.
Each POEMMA telescope employs a 1.6-meter diameter hybrid focal plane, the POEMMA Fluorescence Camera (PFC) comprises the majority of the area while the POEMMA Cherenkov Camera (PCC) takes up the rest, as shown in Fig.~\ref{POEMMAfp}.  The PFC is optimized for the measurement of UHECR EAS longitudinal air fluorescence evolution while the PCC is optimized for the measurement of the fast, Cherenkov signals generated by upward-moving and over-the-limb EAS observations. The PFC uses 55 Photo Detector Modules (PDMs) based on the system developed for the JEM-EUSO instrument  \cite{2015ExA....40...19A} and flown in a number of balloon flights \cite{2017ICRC...35..445B, 2019NIMPA.936..237O}. Each PDM consists of 36 64-channel MAPMTs, with a BG3 filter located on each MAPMT to constrain the wavelength to that of the UV fluorescence band (300$-$500 nm) to minimize the effects of the atmospheric dark-sky airglow background. The PFC design consists of 26,720 $3\times3$ mm$^2$ pixels and will record signals using 1 $\mu$sec temporal sampling to measure the waxing and waning development of the UHECR-induced EAS over time frames of $10 - 100+~\mu$s .
The smaller area PCC consists of 30 focal surface units (FSUs) with each comprised of a 512-channel array of silicon photomultipliers (SiPMs), whose broader wavelength response, 300$-$1000 nm, is better matched to inherent spectral variability of the optical Cherenkov light measurement due to atmospheric attenuation effects. 
When the POEMMA telescopes are pointed towards the Earth's limb, a region corresponding to $9^\circ$ in elevation angle (defined from the nadir direction) and approximately $30^\circ$ in azimuth is observed for the Cherenkov signals, including searching from upward-moving EAS sourced from tau neutrino interactions in the Earth.  This angular mapping on the focal plane is illustrated by the SiPM (red) component in Fig.~\ref{POEMMAfp}.  The POEMMA PCC leads to a modest $\sim10\%$ reduction in the UHECR instantaneous geometry factor for the PFC for fluorescence UHECR observations.  In stereo fluorescence mode, POEMMA will have remarkable sensitivity to UHE neutrinos above 20 EeV, due to the precise angular, energy, and EAS profile measurements required for the UHECR spectrum and composition measurements.  It should be noted that a comparison of the air fluorescence response of SiPMs with significant PDE in the near UV demonstrates that, at least from a photo-detection standpoint, currently available SiPMs have nearly identical response as MAPMTs, but each have specific UV filter requirements due the much larger wavelength bandpass of SiPMs and that of the dark-sky airglow background \cite{2021NIMPA.98564614K}. Thus the potential exists for space-based SiPM-based instruments tuned to the EAS air fluorescence signal with less massive focal planes and without the need for high voltage. 

\section{POEMMA UHECR Measurement Performance and UHECR Science}

Fig.~\ref{POEMMAexposure} illustrates the gains in exposure using space-based UHECR measurements in the context of POEMMA.  Assuming 5-year of POEMMA-Stereo operation, the total exposure is expected to be $\sim 8 \times 10^5$  km$^2$ ster years with precision measurements of UHECRs above 40 EeV: energy resolution of $< 20$\%, an angular
resolution of $\le 1.5^\circ$ above 40 EeV; and a \xmax~ resolution of $\le 30$ g/cm$^2$, which allows for the identification of proton, helium, nitrogen, and iron in a mixed UHECR composition \cite{krizmanic2013modeling}. The left figure shows the exposure growth  as a function of time and comparing this to the growth anticipated from PAO and the TAx4 upgrade \cite{Kido:2019enj}.  The right plot shows the anticipated 5-year POEMMA exposure in context of that reported by PAO (right y-axis scale) and TA (left y-axis scale) at the 2019 ICRC. Above 40 EeV, the yearly UHECR exposure of POEMMA-Stereo is more than
4 times higher than that of the PAO ground array, and 18 times higher compared to the
TA ground array. Above 100 EeV, the POEMMA gain in exposure increases nearly twofold for each comparison.  It should be noted that while ground-arrays operate with nearly  100\% duty cycle, the precision of the measurements is not yet at the level provided by the stereo-fluorescence technique, although improvements in electron/muon identification and radio measurements form the EAS could significantly improve the precision of ground-array EAS measurements \cite{AugerPrime}.  Currently, the UHECR fluorescence measurements performed by PAO and TA have $\sim 10$\% duty cycle, thus the ground-based fluorescence exposure is $\sim 10$\% of that shown in left plot in Fig.~\ref{POEMMAexposure}.  If POEMMA performs UHECR measurements in tilted, POEMMA-Limb mode, Fig.~\ref{POEMMAexposure} shows the significant gain in exposure, but at a cost of increased UHECR detection energy threshold and reduced precision on the EAS measurements.
Conservatively assuming two monocular
EAS measurements, simulation of the POEMMA-Limb UHECR response above 40 EeV yield an energy resolution of $\sim$30\%, an angular resolution of $< 10^\circ$, and an Xmax resolution of $\sim$ 100 g/cm$^2$, which is sufficient to distinguish between proton and iron UHECR primaries \cite{2020PhRvD.101b3012A}.  While the POEMMA spacecraft are planned to be used in tandem, the ability to perform monocular measurements is needed for mission risk mitigation.

\begin{figure}[ht]
\begin{minipage}[t]{0.49\textwidth}
  \includegraphics[width=0.99\textwidth]{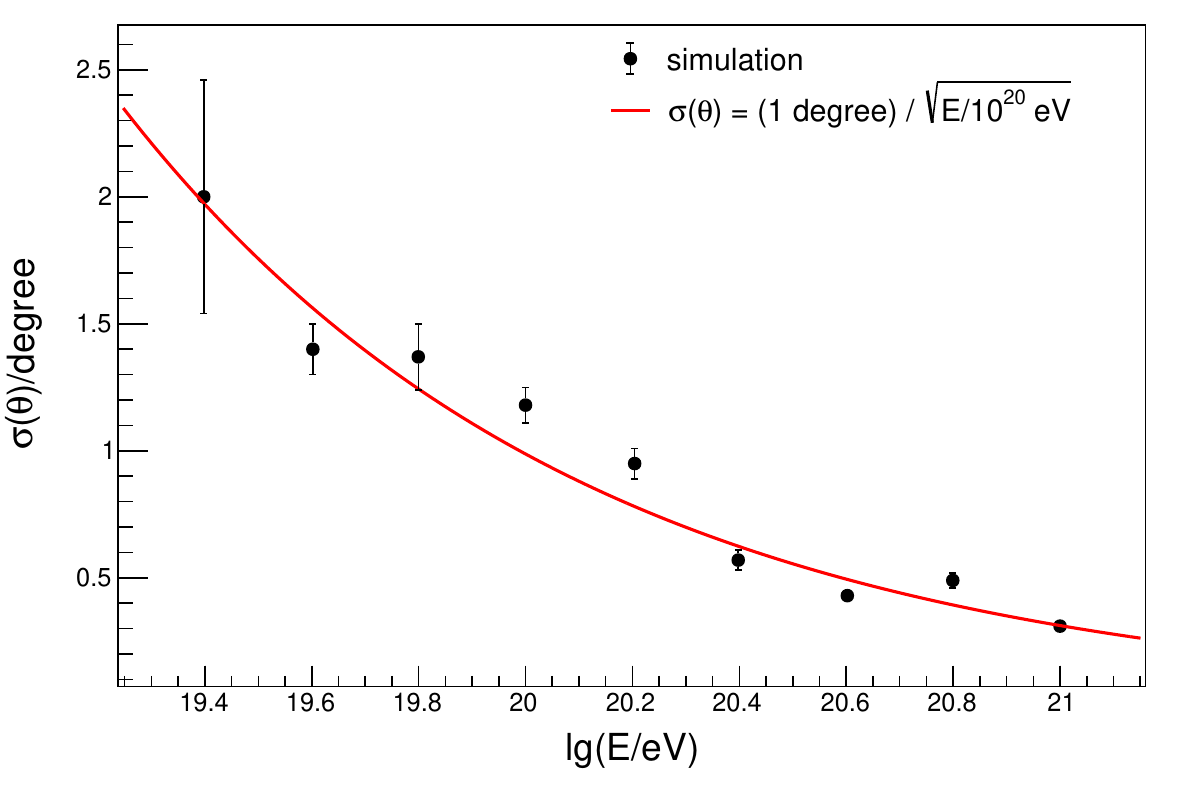}
\end{minipage}
\hfill \begin{minipage}[t]{0.49\textwidth}
   \includegraphics[width=0.99\textwidth]{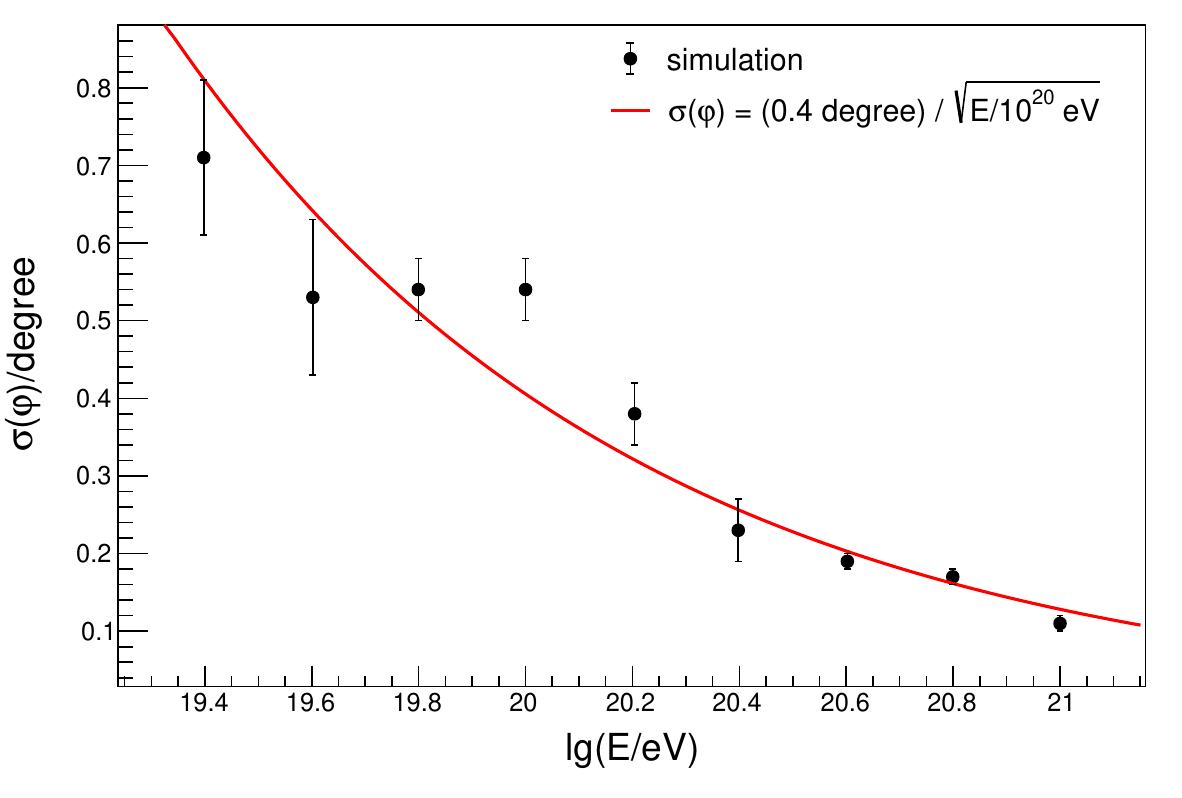}
\end{minipage}
    \caption{The UHECR EAS simulated stereo-reconstructed angular resolution versus
UHECR energy for POEMMA: left: azimuth, right: zenith.
From Ref. \citenum{2020PhRvD.101b3012A}.
      DOI:\href{https://doi.org/10.1103/PhysRevD.101.023012}{10.1103/PhysRevD.101.023012}}
\label{POEMMAangRes}
\end{figure}

\begin{figure}[t]
\begin{center}
    \includegraphics[width=0.99\textwidth]{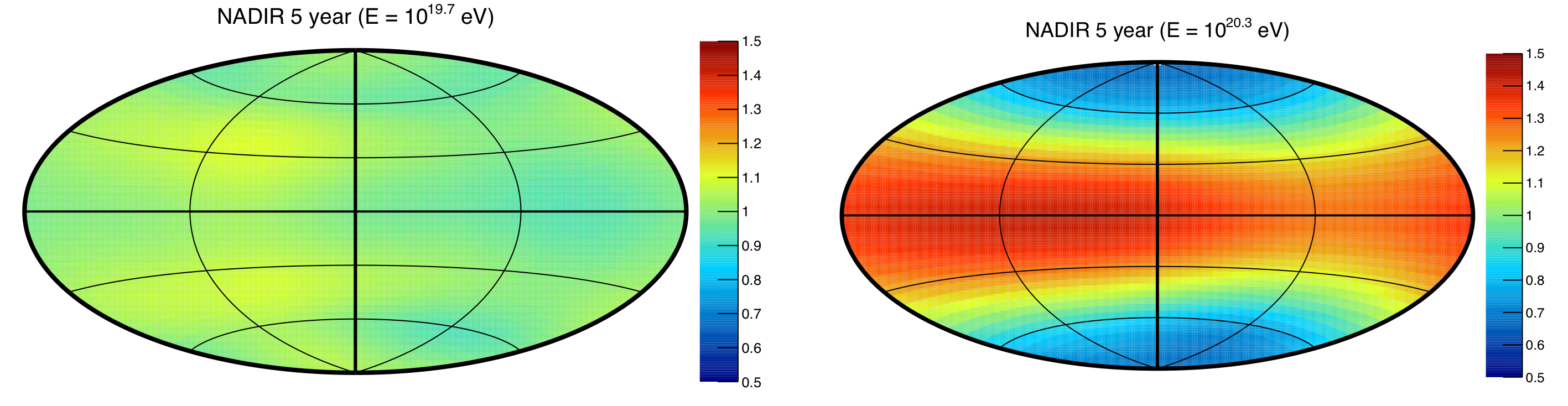}
 \end{center}
\vspace{-0.3cm}
  \caption{The sky exposure for POEMMA-Stereo UHECR observations in declination versus
right ascension. The Color scale denotes the exposure variations
accounting for the sun and moon positions during a 5-year mission. Left: sky exposure
for UHECRs of 50 EeV Right: sky exposure for showers of 200 EeV.
Ref.~\citenum{2020JCAP...03..033K}.
  DOI: \href{https://doi.org/10.1088/1475-7516/2020/03/033}{10.1088/1475-7516/2020/03/033}
}
\vspace{-3 mm}
  \label{POEMMAsky}
\end{figure}

\begin{figure}[t]
\begin{center}
    \includegraphics[width=0.99\textwidth]{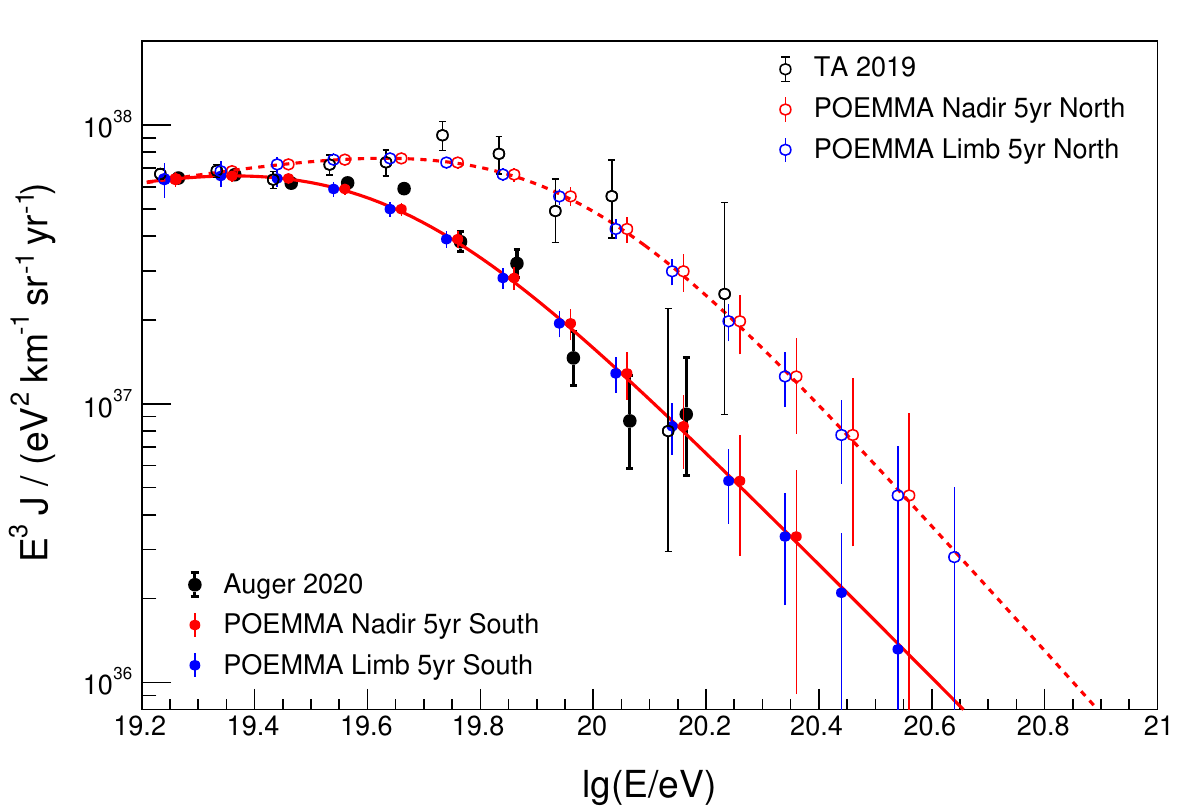}
 \end{center}
\vspace{-0.3cm}
  \caption{The simulated POEMMA spectra extrapolated from and compared to the PAO 2020 spectrum (black dots and
solid line) \citenum{PhysRevLett.125.121106} and the extrapolation and comparison tp the TA 2019 spectrum (black open circles and dotted line) from \citenum{Ivanov:20198M}
for the POEMMA-Stereo (red) and POEMMA-Limb (blue) observational modes, for UHECRs above 16 EeV.
Ref.~\citenum{2020JCAP...03..033K}.
  DOI: \href{https://doi.org/10.1088/1475-7516/2020/03/033}{10.1088/1475-7516/2020/03/033}
}
\vspace{-3 mm}
  \label{POEMMAspectra}
\end{figure}

\begin{figure}[ht]
    \includegraphics[width=0.99\textwidth]{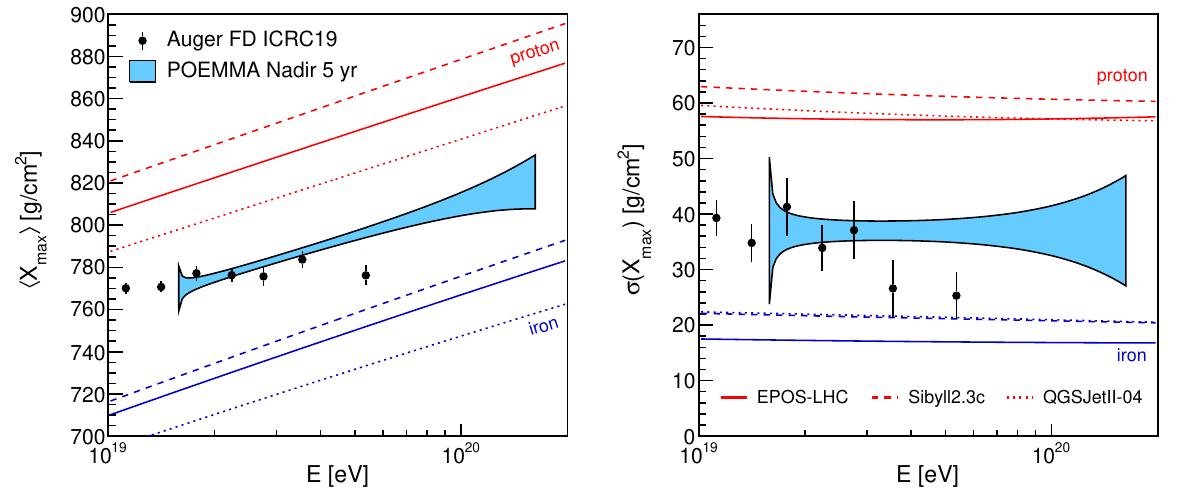}
\caption{The simulated capability of POEMMA to measure $\langle X_\text{max}\rangle$
  and $\sigma(X_\text{max})$ for UHECR composition studies. A simple UHECR composition model based on lower energy PAO measurements was used to assess the performance. The blue band maps the statistical uncertainties in five
  years of POEMMA-Stereo measurements versus number of events per 0.1 in the logarithm of energy, the $X_\text{max}$ resolution and efficiency for $\theta < 70^\circ$, and an intrinsic  EAS-to-EAS fluctuations of 40~g/cm$^2$ was assumed. The black dots show the PAO  fluorescence UHECR data presented at the 2019 ICRC~\cite{Yushkov:2020nhr}.
Ref.~\citenum{2020JCAP...03..033K}.
  DOI: \href{https://doi.org/10.1088/1475-7516/2020/03/033}{10.1088/1475-7516/2020/03/033}}
\label{POEMMAXmax}
\end{figure}

As with ground-based UHECR experiments, space-based experiments rely on detailed Monte Carlo simulations to predict the accuracy and science return of the measurements.  Here, those developed for POEMMA are summarized. Fig.~\ref{POEMMAangRes} shows the stereo reconstructed angular resolution as a function of simulated UHECR energy, where the telescopes are separated by 300 km and tilted to view a common volume along the orbit path.  The strength of the stereo reconstruction technique is evident as both the zenith and azimuth angular resolution are $\sim 1.5^\circ$ for $E_{CR} \ge 40$ EeV.  This angular resolution with the stereo technique yields simulated UHECR energy resolution $<20\%$ for $E_{CR} \ge 50$ EeV \cite{2020PhRvD.101b3012A}. The UHECR sky coverage for these UHECR observations from a 525 km altitude and 28.5$^\circ$ inclination orbit are shown in Fig.~\ref{POEMMAsky} assuming 5-years of observations at 50 and 200 EeV. The results demonstrate full sky-coverage and $\pm 5$\% variation at 50 EeV and $\pm 50$\% variation at 200 EeV. 

The ground-based PAO and TA UHECR spectrum measurements demonstrate a significant difference, possibly indicating the UHECR sources are different in the northern and southern hemispheres.  Assuming the PAO and TA UHECR spectral parameters reported at the 2019 ICRC and POEMMA's simulated response, Fig.~\ref{POEMMAspectra} shows the predicted measurements assuming 5-years of both stereo and limb (tilted) observations, showing the possible extension well past 100 EeV.  This capability allows for the search of spectral recovery at the highest energies \cite{2020PhRvD.101b3012A} due to the $10^{6}$ km$^2$ sr year scale exposures provided by 5 years of space-based UHECR observation.  

The stereo reconstruction technique combined with fine sampling of the EAS development evolution provide a \xmax~ resolution of $\lsim 30$ g/cm$^2$ that improves to $< 20$ g/cm$^2$ above 100 EeV, and thus provides accurate determination of the UHECR nuclear composition \cite{2020PhRvD.101b3012A}. Fig.~\ref{POEMMAXmax} presents this capability in terms of extending to higher energy a model based on PAO lower energy measurements.  The results show good determination above 100 EeV, well before statistical errors begin to dominate the results.

\begin{figure}[ht]
\begin{center}
    \includegraphics[width=0.32\textwidth]{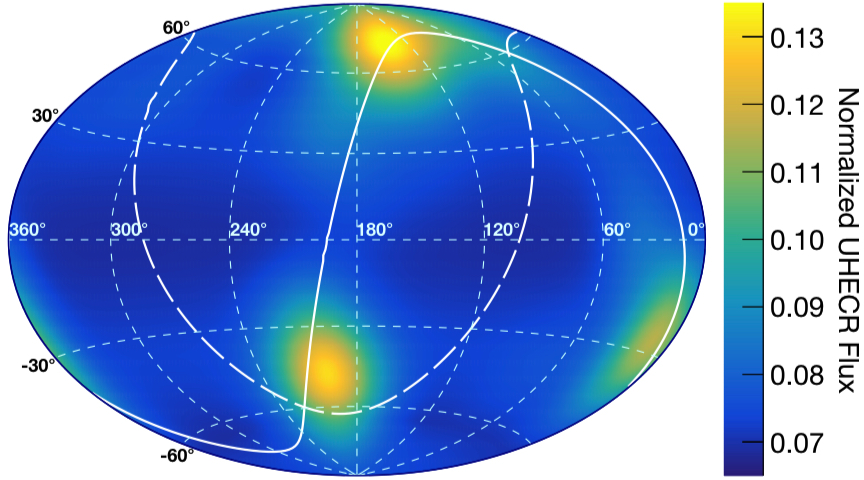}
    \includegraphics[width=0.32\textwidth]{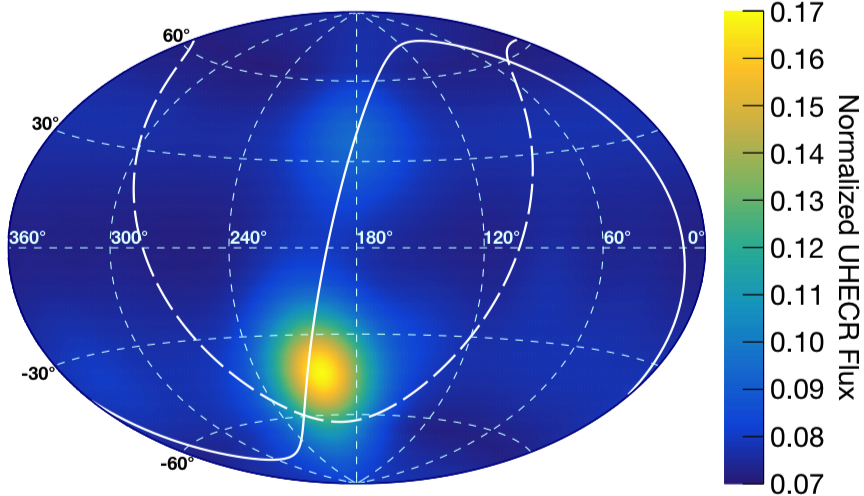}
    \includegraphics[width=0.32\textwidth]{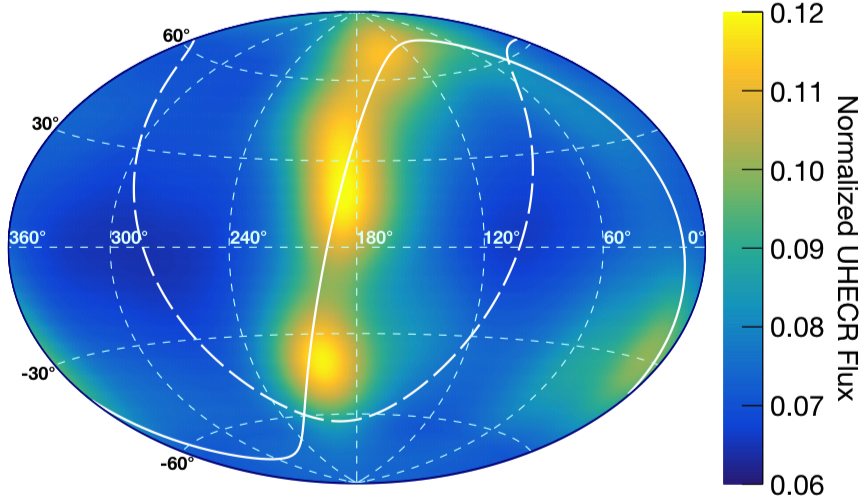}
 \end{center}
\vspace{-0.3cm}
  \caption{The equatorial coordinate sky maps of simulated POEMMA UHECR measurements for different astrophysical catalogs using the best fit parameters reported by the PAO collaboration \cite{PierreAuger:2019phh} Left: starburst galaxies with 11\% anisotropy fraction; Middle: {\it Swift}-BAT AGNs with 8\% anisotropy fraction; Right: 2MRS catalog with 19\% anisotropy fraction. All assume 15\% angular spread of the arrival directions. From Ref. Ref.~\citenum{2020JCAP...03..033K}.
  DOI: \href{https://doi.org/10.1088/1475-7516/2020/03/033}{10.1088/1475-7516/2020/03/033} \label{POMMAsbMAPS}}
\vspace{-3 mm}
\end{figure}

In 5-year operation, POEMMA is anticipated to measure over 1,400 UHECRs above 40 EeV with high precision given the good measurement resolutions.  Fig.~\ref{POMMAsbMAPS} shows the source sky maps obtained by three different assumptions regarding the astrophysical distribution of UHECRs \citenum{2020JCAP...03..033K} based on the PAO results and correlation analysis with similar catalogs \cite{PierreAuger:2019phh}. The different results in the upper and lower hemispheres predicted by these demonstrate why full sky coverage with precision UHECR measurements is important for definitively identifying the astrophysical sources of UHECRs.

It is noted that the UHECR measurement performance shown through modeling and simulations also expands the search for UHE neutrinos, UHE photons, and measurement of the proton-air cross section at $\sqrt{s} = 450$ TeV \cite{2020PhRvD.101b3012A} while also provided unique sensitivity to the detection of super-heavy dark matter (SHDM) from decay or annihilation into UHE neutrinos \cite{2021PhRvD.104h3002G} or UHE photons \cite{2020PhRvD.101b3012A}.

\subsection{Effects due to the Dark-sky Airglow Background and Cloud Cover}

\begin{wrapfigure}{r}{0.55\columnwidth}
\captionsetup{width=0.5\columnwidth}
%\begin{figure}[h]
\centering
\includegraphics[width=0.5\columnwidth]{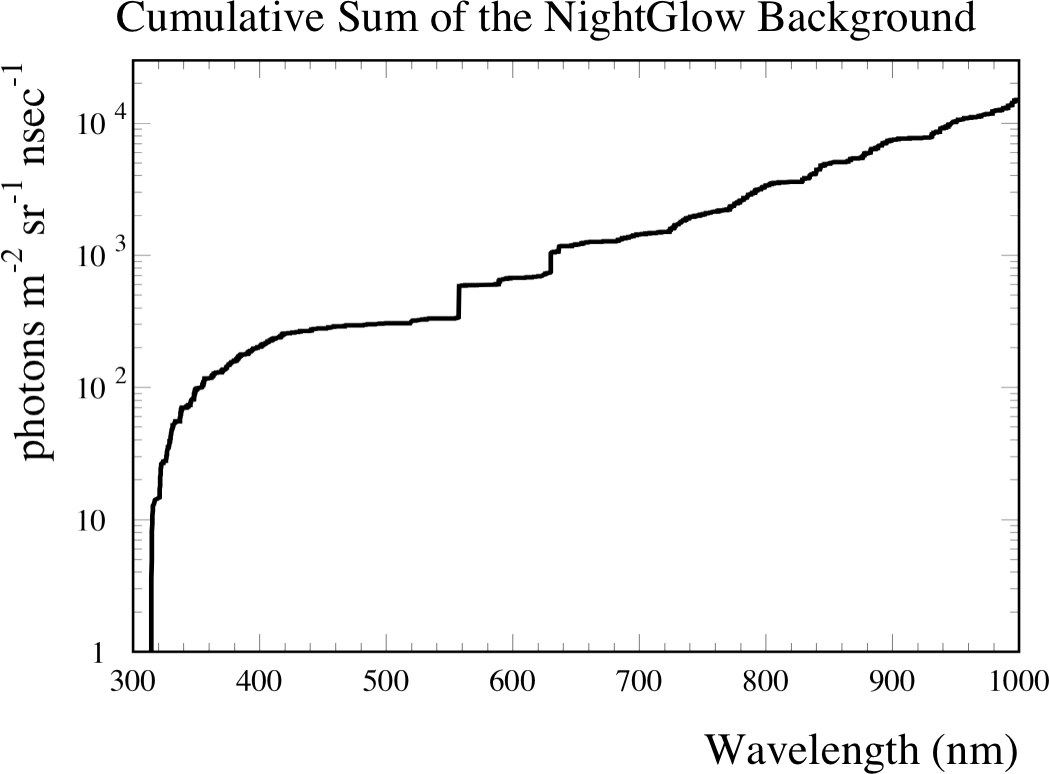}
\caption{The cumulative sum of the dark-sky airglow background using the measurements of Hanuschik \cite{2003A&A...407.1157H,2006JGRA..11112307C}. From Ref.~\citenum{2021NIMPA.98564614K}.
  DOI: \href{https://doi.org/10.1016/j.nima.2020.164614}{10.1016/j.nima.2020.164614}
}
\label{AirGlow}       % Give a unique label
%\end{figure}
\end{wrapfigure}

Optical measurements of EAS are performed with a persistent atmospheric background even during astronomical night. The molecules and atoms in the upper atmosphere of the Earth are continuously ionized by solar and cosmic radiation leading to production of faint light that is commonly known as an airglow \citep{Meier,Leinert}. These chemiluminescent processes are concentrated in a $\sim$10 km thick layer around 95 km in altitude over the entire planet and are different than the geomagnetic processes that cause aurora near the polar regions.  When views from LEO, the dark-sky airglow light together with the reflected component of the starlight and zodiacal light act as a diffuse night atmospheric background that effectively sets the energy threshold for EAS detection. For space-based UHECR EAS measurements, the directly viewed zodiacal light is not an issue, just the atmospheric scattered component. Airglow is a dynamic phenomenon that varies in with geographical position, season, solar activity, geomagnetic activity and changes in the Earth's atmosphere \citep{Shepherd,Deutsch,Pfaf}. Additionally, the altitude of the observatory and its viewing direction in terms of angle away from nadir further yields to an increase in intensity, in part by viewing a longer path thru the dark-sky airglow layer, and is described by the van Rhijn formula \cite{Roach}. Fig.~\ref{AirGlow} presents the dark-sky airglow measurements by Hanuschik near maximal solar activity \cite{2003A&A...407.1157H,2006JGRA..11112307C} for the wavelength band 300$-$1000 nm in units of photons m$^{-2}$  ns$^{-1}$ sr$^{-1}$ as a cumulative sum as a function of wavelength. 
Given the strong wavelength dependence, the impact of dark-sky airglow on the measurements of EAS air fluorescence and optical Cherenkov needs to be considered and an analysis for both is detailed in Ref.~\citenum{2021NIMPA.98564614K}.  Since the main contribution to the air fluorescence is from 300$-$450 nm, near-UV filters combined with the wavelength response of bi-alkali PMTs limits the airglow intensity. A value of $\sim$ 500 photons m$^{-2}$ ns$^{-1}$ $sr^{-1}$ for the dark-sky airglow in the 300$ -$400 nm wavelength band is based on measurements from balloon altitudes by the NIGHTGLOW experiment nadir-viewing measurements \cite{2005APh....22..439B}, slightly more but consistent with the measurements of Hanuschik \cite{2003A&A...407.1157H,2006JGRA..11112307C}.
NIGHTGLOW also measured a van Rhijn type enhancement as a function of viewing angle away from nadir.  A straightforward calculation based on an instrument's effective optical collecting area, iFoV, and signal integration time provides a mean measurement of the airglow background in a pixel of a telescope.  Assuming $A_{Eff}$ = 6 m$^2$, iFoV = $0.084^circ$, and 1 $\mu$s signal integration time, e.g. the POEMMA on-axis response, a airglow background of 500 photons m$^{-2}$ ns$^{-1}$ $sr^{-1}$ yields a per pixel mean photon background of $\approx 6.5$ photons.  Assuming an effective quantum efficiency of 30\% for the PMT channel, this yields a background of $\approx 2$ photo-electrons (PEs).  Thus any EAS fluorescence signal needs to be well-above this to achieve a sufficient signal-to-background ratio to only trigger the instrument on EAS events versus the airglow background.  In practice, any instrument will occasionally operate to trigger on the background to provide a calibration assessment.

\begin{figure}[ht]
\begin{center}
    \includegraphics[width=0.49\textwidth]{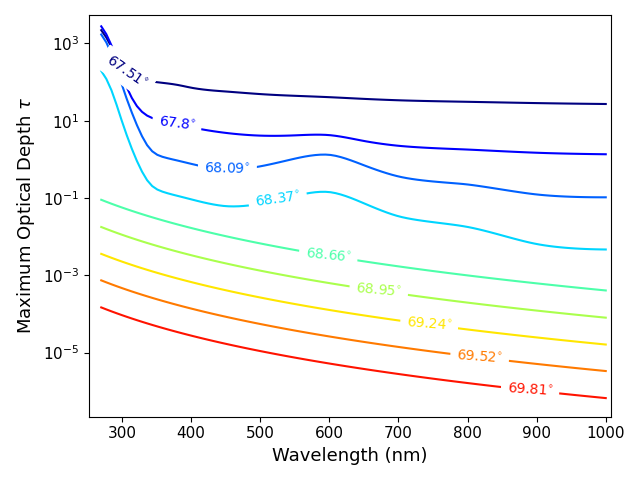}
    \includegraphics[width=0.49\textwidth]{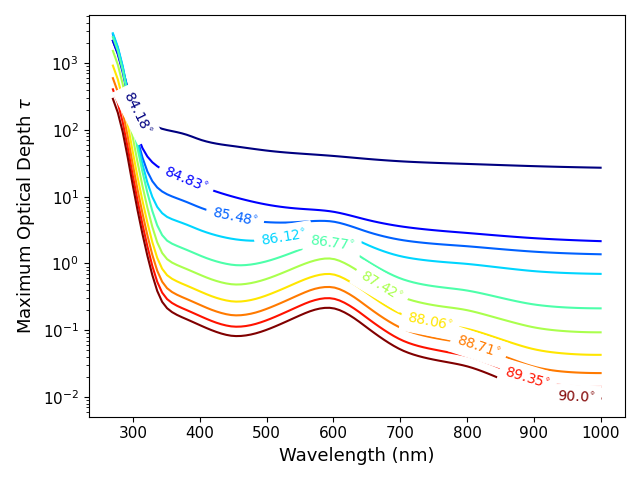}
 \end{center}
\vspace{-0.3cm}
  \caption{The optical depth as a function of wavelength for viewing the EAS optical Cherenkov from over the limb of the Earth VHECRs. The different curves are the optical depths for different angles away from nadir. Left: the optical depth for viewing from 525 km altitude orbit where the limb of the Earth is at an angle 67.51$^\circ$ away from nadir. Right: the optical depth for viewing from 525 km altitude orbit where the limb of the Earth is at an angle 84.18$^\circ$ away from nadir.
Ref.~\citenum{PhysRevD.104.063029}. 
          DOI:\href{https://doi.org/110.1103/PhysRevD.104.063029}{10.1103/PhysRevD.104.063029} \label{VHECRopDepth}
  }
\vspace{-3 mm}
\end{figure}

The shape of the optical Cherenkov spectrum emitted from either upward-moving EAS sourced from tau neutrino interactions in the Earth or from viewing VHECRs above the Earth limb is highly variable over the band starting at 300 nm to above 1000 nm \cite{2021NIMPA.98564614K,2021PhRvD.103d3017C}. This is due to large differences in the wavelength-dependent atmospheric attenuation due the different amounts of aerosols, ozone, and atmosphere itself, that the Cherenkov light from different EAS must propagate through to reach the detector. This is effect is quantified by examining the optical depth as a function of angle above the limb of the Earth\footnote{presented as angle from nadir where angle from the nadir for the limb is defined by the altitude of observation and the radius of the Earth.}, which is shown in Fig.~\ref{VHECRopDepth}. When the VHECR is viewed from a detector at 525 km altitude, for angles $\lsim 1^\circ$ above the limb, the attenuating effects of the atmosphere become quite small.  However, for the 33 km altitude case, a significant column depth of ozone remains that significantly attenuates the Cherenkov signal $\sim 350$ nm \cite{1999ICRC....2..388K}, even out the detector horizontal angle (90$^\circ$ from nadir). Given the extreme variability of the wavelength dependence of the optical depth, the Cherenkov spectrum at the instrument can be peaked throughout the range for $<300-1000$ nm for VHECRs and upward-moving Earth-emergence \taon EAS \cite{2019PhRvD..99f3011R,2020PhRvD.102l3013V,2021PhRvD.103d3017C,PhysRevD.104.063029}. This motivates the use of SiPMs as the photo-detectors since they have a response matched to the optical Cherenkov wavelength band, including a SiPM variety with peak response around 450 nm \cite{OTTE2017106}. Over the entire wavelength band shown in Fig.~\ref{AirGlow}, the total count rate is $\sim$ 15,000 photons m$^{-2}$ ns$^{-1}$ $sr^{-1}$, but the fast, $\sim 10$ ns time spread of the optical Cherenkov signals when viewed a few degrees from the EAS trajectory \cite{PattersonHillas, 2019PhRvD..99f3011R,2021PhRvD.103d3017C} helps to reduce the effects of the airglow background. Choosing SiPMs with  blue-peaked response also have photo-detection-efficiencies (PDE) $\lsim 20\%$ above 700 nm, which also helps to reduce the airglow background. For $A_{Eff}= 1$ m$^2$, iFoV$=0.1^\circ$, $\Delta t = 10$ ns, and an effective PDE$= 30\%$ over the 200$-$1000 nm band yields a mean background of $\approx 0.14$ PEs.

\begin{figure}[ht]
\begin{center}
    \includegraphics[width=0.42\textwidth]{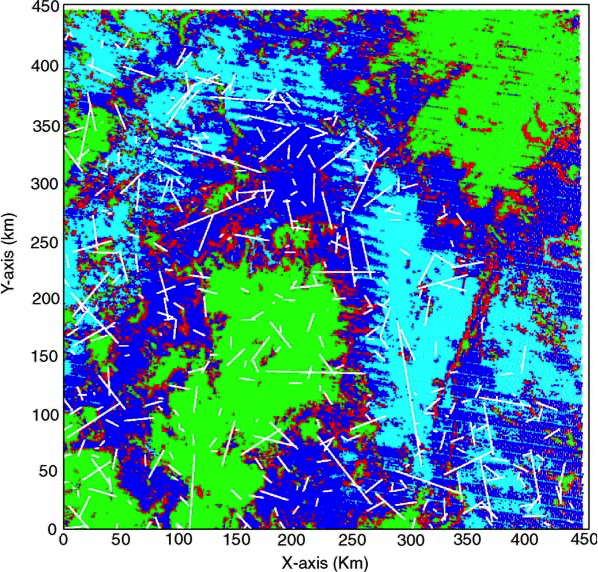}
     \includegraphics[width=0.54\textwidth]{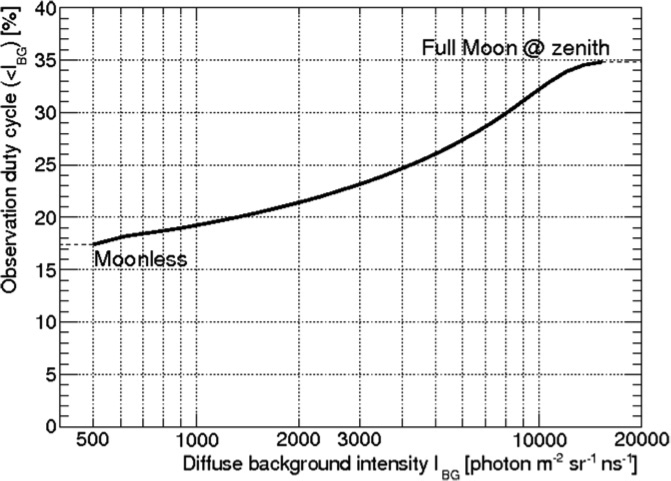}
 \end{center}
\vspace{-0.3cm}
  \caption{Left: The projected 100 EeV simulated observed UHECR EAS tracks (white lines), which trigger the OWL instrument, superimposed over measured MODIS cloud cover, in approximately one quadrant of the square defined by the embedded quasi- circular OWL viewed ground area. Key: (Light blue) high-confidence cloud free; (dark blue) cloud free; (red) probably cloudy; (green) cloudy. For this particular cloud frame,  20\% of the simulated track sample have a ‘‘Clear’’ aperture as defined by no cloudy or probably cloudy MODIS pixels within 3 km of a track. From the MODIS database used in this study, $\sim 50\%$ of the cloud heights were $\le 4 km$ and below the visible portion of the EAS \cite{2003ICRC....2..639K}. From Ref. \citenum{2004APh....20..391S}.
  DOIs: \href{https://doi.org/10.1016/S0927-6505(03)00196-8}{10.1016/S0927-6505(03)00196-8}. Right: The night-sky observational duty cycle ($\eta$) as a function of astronomical dark-sky and moonlight induced background as a function of moon phase.
  From Ref. \citenum{2013APh....44...76A}.
  DOIs: \href{https://doi.org/10.1016/j.astropartphys.2013.01.008}{10.1016/j.astropartphys.2013.01.008}. \label{CloudDC}
  }
\vspace{-3 mm}
\end{figure}

Since space-based instruments view different atmospheric conditions as the instruments orbit the Earth, the distribution of clouds are inherently  variable
and data from Earth-observing satellites together with UHECR simulations provide a mechanism to assess the impact \cite{2004APh....20..391S,2003ICRC....2..639K,2013APh....44...76A,2013APh....44...76A}.  The left panel in Fig.~\ref{CloudDC} shows the distribution of observed 100 EeV UHECR EAS simulated for the OWL mission overlaid on a particular frame of MODIS cloud measurements \cite{2004APh....20..391S}.  When cloud height measurements are included and using long-duration cloud measurement data sets, they predict between 50$-$60\% of the observable portion of the simulated EAS are above the clouds. Furthermore, if the location of EAS signal maximum occurs above the cloud, a high fraction of events that enter clouds can be accurately reconstructed, especially if the precise spatial location of the EAS is provided by stereo fluorescence measurements \cite{Abu-Zayyad:2003lhv}. 

The level of scattered moonlight also affects the duty cycle of UHECR observations, and the level in terms of the units of dark-sky background versus moon phase is shown in the right panel of Fig.~\ref{CloudDC}. Practically, accepting more moonlight for UHECR observations effectively raises the detection energy threshold due to the higher moonlight (plus airglow) background.  Flight dynamic studies performed for the JEM-EUSO ISS implementation (400 km altitude, 51.6$^circ$ inclination) \cite{2003ICRC....2..639K}and POEMMA (525 km altitude, 28.5$^circ$ inclination) \cite{2020PhRvD.101b3012A} calculate that $\sim$20\% of the time, the spacecraft orbit will be during astronomical night with minimal effects from moonlight.  Assuming that 60\% of UHECR EAS are observable, a 12\% total duty cycle is calculated for UHECR experiments in LEO.

\section{UHECR EAS Optical Measurements using Balloon-borne Instruments}

While orbiting neutrino-detection missions using the EAS optical signals can be performed in multi-year mission time frames, the availability of ultra-long duration balloon (ULDB) flights, potentially of 100-day duration allows for meaningful UHECR measurements in a relatively cost-effective suborbital space flight platform. These $\sim$30 km altitude flights are launched from latitudes, e.g. Wanaka, New Zealand, that allow for astronomical night operation, provide an environment to view the  air fluorescence signal of downward-moving EAS from UHECRs as well as the optical Cherenkov signal above-the-limb VHECRs while search for from neutrino-induced, Earth-emergent upward-moving EAS. The EUSO-SB2 ULDB experiment, scheduled to launch in 2023, consists of a downward-looking fluorescence telescope (FT) to record the development of the fluorescence EAS signals from UHECRs   \cite{Eser:2021H6,Osteria:2021OJ,Filippatos:2021b5}. The FT \cite{Osteria:2021OJ} consists of a $\sim$1-meter diameter optical aperture Schmidt telescope with a $37.4^\circ \times 11.4^\circ$ FoV. The focal plane of the 6912 pixels in the FT are contained in three Photo Detection Modules (PMDs) developed for the EUSO mission \cite{2017NIMPA.866..150A}, which are planned for use in the POEMMA mission. Each PDM consisting of 36 Multi-Anode Photomultipler tubes (MAPMTs) and SPACIROC3 front-end electronic (FEE) readout that employ 1 $\mu$s sampling of the UHECR EAS development and has a channel to record fast, $\gsim 10$  ns optical signals such as that from Cherenkov reflection off the ground. This PDM design has been flown in the ULDB EUSO-SB1 mission in 2017 \cite{2019NIMPA.936..237O} and EUSO-Balloon test flight mission \cite{2017ICRC...35..445B} flown in 2014, and Mini-EUSO instrument \cite{2021ApJS..253...36B} operating on the International Space Station starting in 2019. This PDM development path illustrates one of the benefits of using relatively inexpensive balloon-borne instruments to develop technologies for eventual space-based, orbiting missions. Simulation studies have predicted the UHECR observation rate for EUSO-SPB2 for UHECR fluorescence events is about 1 in 8 hours for events with energies slightly above 1 EeV, assuming clear observation during dark moonless nights \cite{Filippatos:2021b5}, which reflects the inherent difficulty of performing a UHECR air fluorescence experiment even from a 100-day ULDB balloon mission.

\begin{figure}[ht]
\begin{center}
    \includegraphics[width=0.36\textwidth]{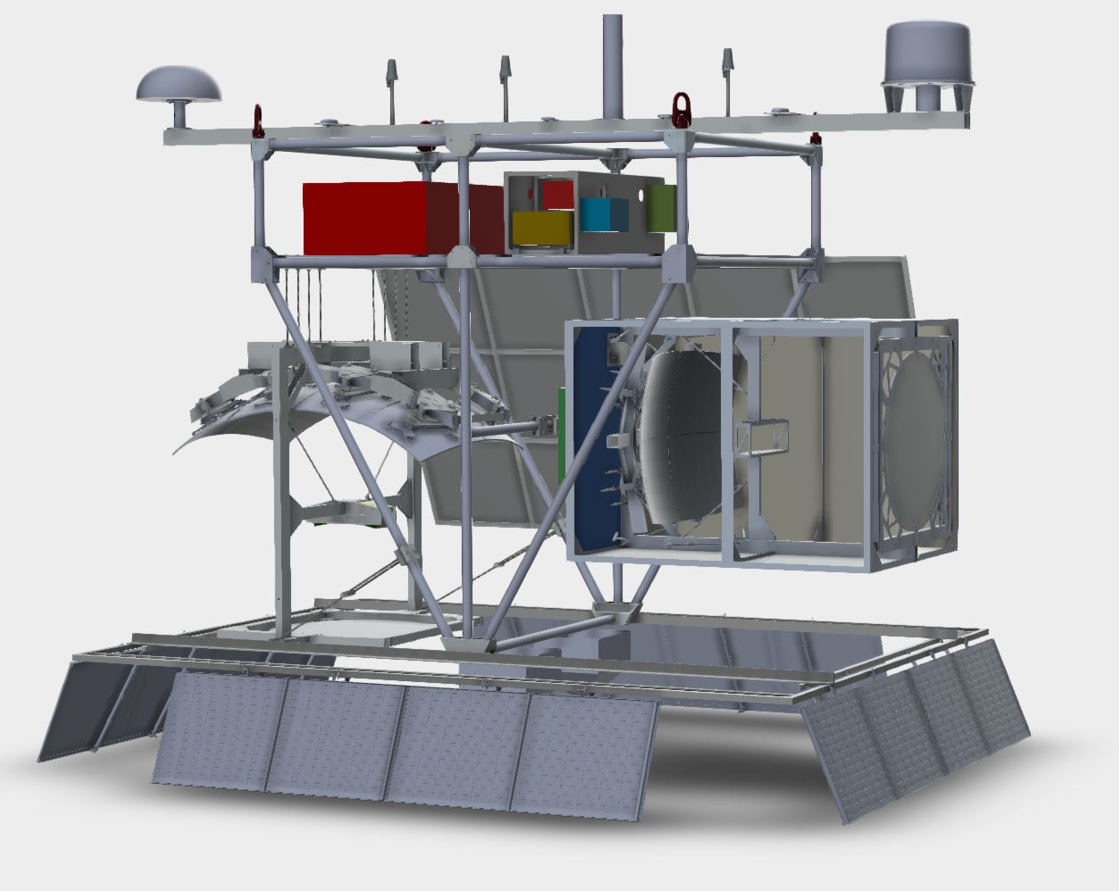}
    \includegraphics[width=0.32\textwidth]{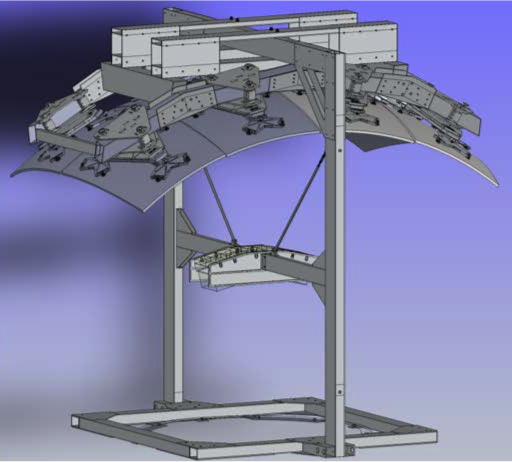}
    \includegraphics[width=0.29\textwidth]{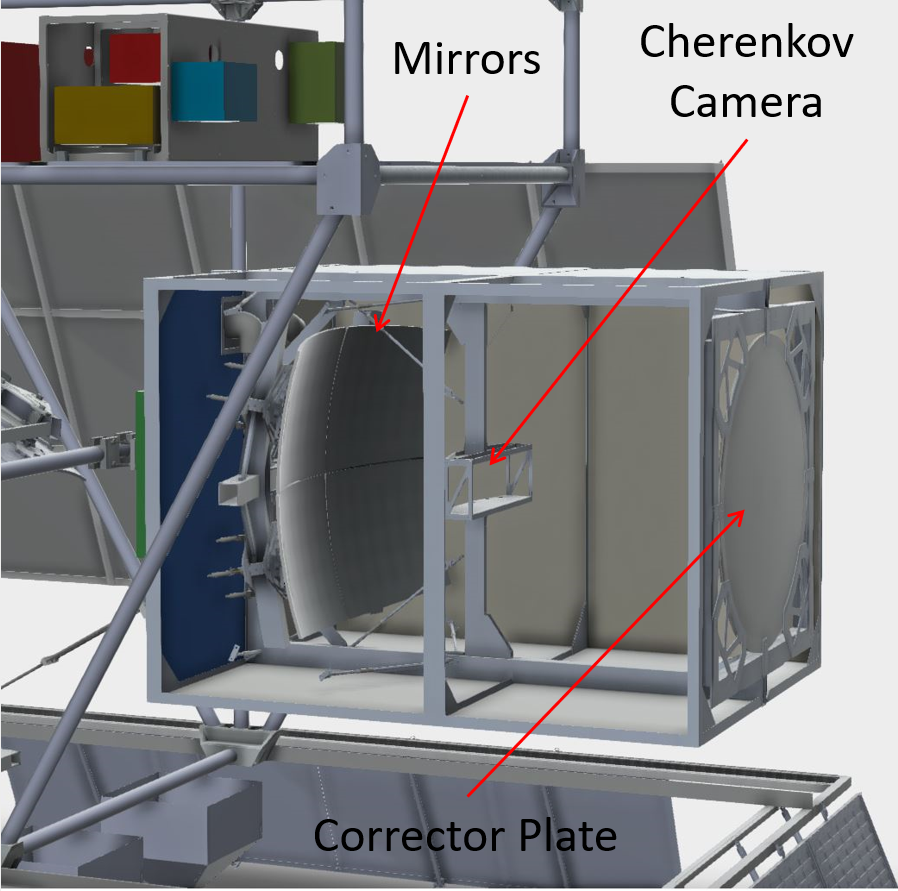}
 \end{center}
\vspace{-0.3cm}
  \caption{The EUSO-SPB2 payload configuration. Left: Schematic of the payload. Middle: Schematic of the fluorescence telescope. Right: Schematic of the Cherenkov telescopes. Not shown is the infrared cloud camera (see Ref.~\citenum{Diesing:2021fB} or the airglow monitoring sensor package (see Ref.~\citenum{2019NIMPA.922..150M}). From Refs. \citenum{Osteria:2021OJ,Filippatos:2021b5}.
  DOIs: \href{https://doi.org/10.22323/1.395.0206}{10.22323/1.395.0206}, \href{https://doi.org/10.22323/1.395.0405}{10.22323/1.395.0405}
  }
\vspace{-3 mm}
  \label{SPB2schematics}
\end{figure}

The EUSO-SPB2 payload includes a Cherenkov telescope (CT) \cite{Eser:2021H6,Bagheri:2021s0}, a Schmidt telescope with a $12.8^\circ \times 6.4^\circ$ horizontal by vertical (above or below the viewed Earth limb) FoV. The schematic view of the EUSO-SPB2 payload is shown in Fig.~\ref{SPB2schematics}.  The focal plane is comprised of a 512 pixel silicon photomultiplier (SiPM) camera using the MUSIC ASIC \cite{10.1117/12.2231095} for FEE readout using a 100 MegaSample/second digitization. The choice of the SiPM is such that it has relatively high sensitivity in the 200$-$1000 nm wavelength band needed due to the inherent variability of the Cherenkov signal \cite{2021NIMPA.98564614K}.  The 10 ns digitization span is optimized for the fast Cherenkov signal that is produced within $\sim$1.5$^\circ$ from the EAS trajectory. The CT optics have a relatively large pixel, or instantaneous, iFoV $=0.4^\circ$, implying the count rate from the dark-sky background will be significant and dominate over the background due to SiPM dark-count rates \cite{Bagheri:2021s0}. One other novel feature of the CT is that a bifocal mirror splits the $0.785$ m$^2$ effective area into two separate images on the focal plane. This provides a coincidence trigger using spatially separated events, but at a cost of modestly increasing the effects of the dark-sky background. The goal of the bifocal design is to determine whether event backgrounds due to ionization in the SiPM pixels, i.e. due to traversing cosmic rays or other radiation backgrounds, require the need of a such a spatially separated trigger. 

\begin{figure}[h]
\begin{center}
    \includegraphics[width=0.99\textwidth]{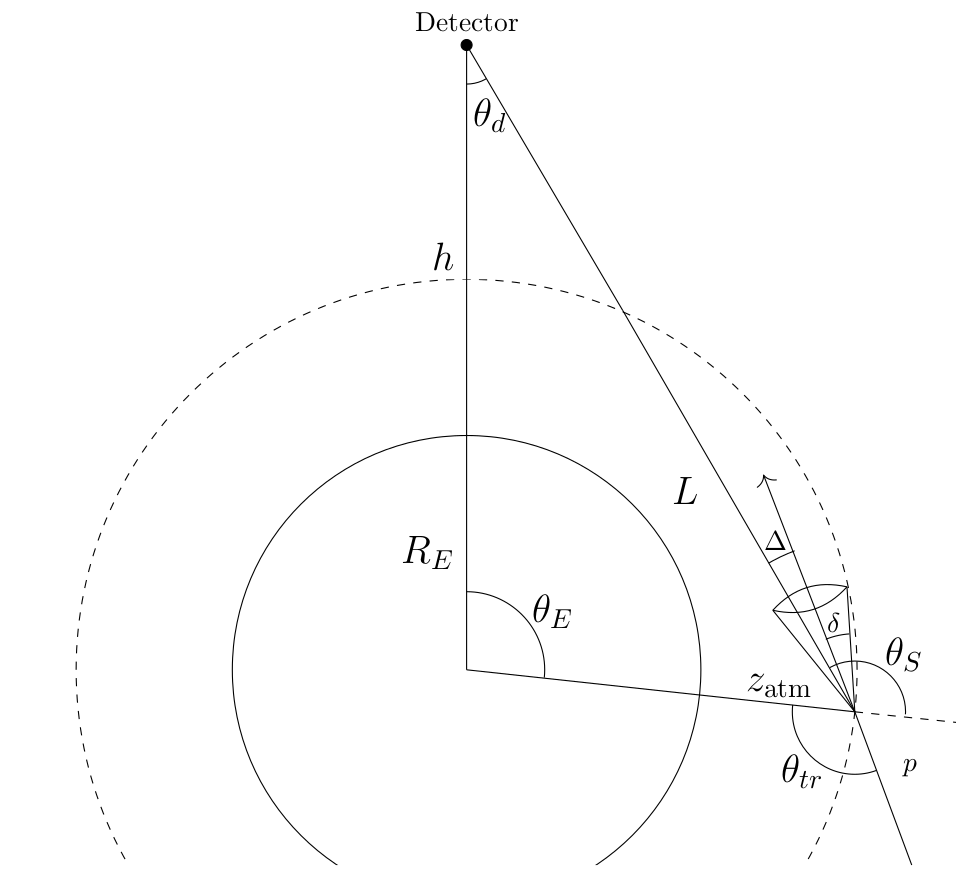}
 \end{center}
\vspace{-0.3cm}
  \caption{The geometry for observing the beamed optical Cherekov signal generated by over-the-limb VHECR EAS in the atmosphere.  From Ref.~\citenum{PhysRevD.104.063029}. 
          DOI:\href{https://doi.org/110.1103/PhysRevD.104.063029}{10.1103/PhysRevD.104.063029}
}
\vspace{-3 mm}
  \label{VHECR}
\end{figure}

\begin{figure}[h]
\begin{center}
    \includegraphics[width=0.49\textwidth]{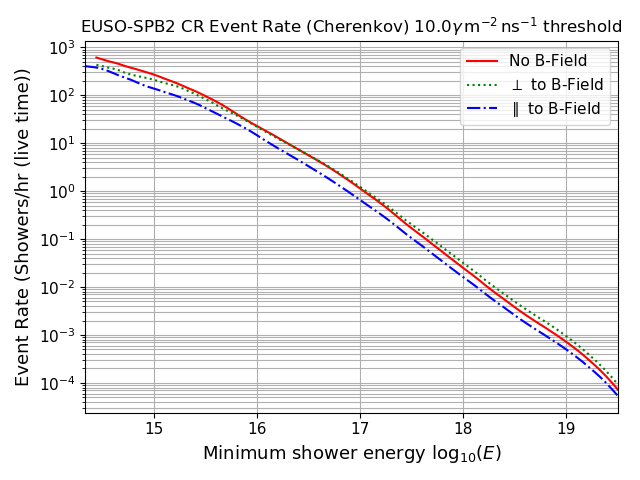}
       \includegraphics[width=0.49\textwidth]{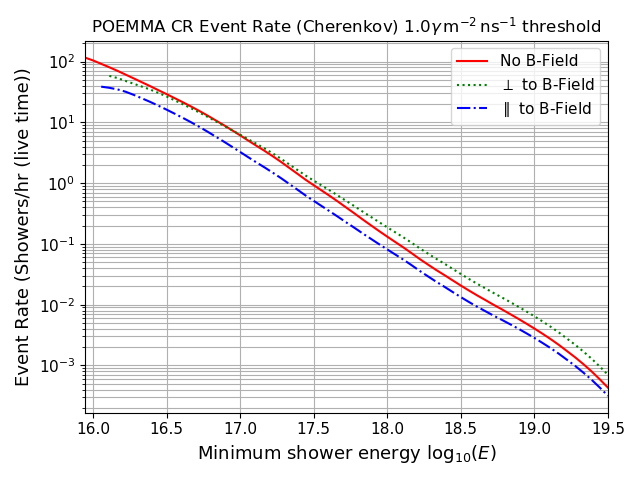}
 \end{center}
\vspace{-0.3cm}
  \caption{The simulated integral cosmic ray event rates (above a threshold $E$) for observing over-the-limb events via the optical Chernekov signal for EUSO-SPB2 (left) and POEMMA (right). The effects of the geomagnetic field are provided by assuming the acceptance for three different conditions, no $\vec{B}$-field, $\vec{B}$-field parallel to the EAS trajectory, and $\vec{B}$-field perpendicular to the trajectory. From Ref.~\citenum{PhysRevD.104.063029}. 
          DOI:\href{https://doi.org/110.1103/PhysRevD.104.063029}{10.1103/PhysRevD.104.063029} \label{VHECRrate}
}
\end{figure}

Fig.~\ref{VHECR} shows the geometry of the VHECR events that are detected in the direction above the limb in relation to the optical Cherenkov EAS signal.  Simulations \cite{PhysRevD.104.063029} have shown that the integral count rates for these events is substantial, and
Fig.~\ref{VHECRrate} presents these rates above a VHECR threshold energy for the EUSO-SPB2 and POEMMA missions.  The results show that the EUSO-SPB2 ULDB experiment should have a VHECR threshold energy well below 1 PeV with a rate of several hundred observed events per hour of live time above threshold.  The rates based are three different assumptions on the geomagnetic field to estimate the possible effects. In practice, setting different analysis thresholds on the Cherenkov signal strength should map out the VHECR rate since the cosmic ray spectrum in this energy range is experimentally fairly well-measured. The simulated rates for POEMMA are also shown in Fig.~\ref{VHECRrate}. These over-the-limb VHECR events do form a background for upward-moving EAS neutrino searches due to the $\lsim 1^\circ$ atmospheric refraction of the VHECR optical Cherenkov signal when viewed near the limb \cite{2020PhRvD.102l3013V}. Thus this measurement provides a measure of this potential background to upward, Earth emerging events from tau neutrino interactions while also providing a course calibration signal for the Cherenkov telescope.

The simulated results presented in Fig.~\ref{VHECRrate} are based on simulating the electromagnetic properties of the EAS, thus ignoring the $\sim$10\% muon contribution.  Since muons are relatively long-lived in terms of EAS age, they could provide an additional Cherenkov signature \cite{2016PhRvD..94l3018N}. If there is a way to use the differences in muon vs electromagnetic EAS Cherenkov signatures, which are strongly dependent on atmospheric attenuation, i.e. viewing angle from the instrument, this would allow a measure of the hadronic content in the detected EAS, 

However, a single measurement within the Cherenkov light pool does not precisely constrain the actual geometry of the observed EAS.  This translates into a more coarse resolution for both the UHECR energy and location of \xmax.
However, the concept of using SmallSat constellations for a variety of space-based measurements is currently being implemented \cite{SmallSatConst}. In principle, a constellation of modest, meter-diameter Cherenkov telescopes could provide enough sampling of the Chernekov light and temporal pool \cite{PattersonHillas, Hillas1,Hillas2} to more precisely measure the properties of the VHECRs, such as is done in ground-based experiments such as the non-imaging Chernekov array at the Bolivian Air Shower Joint Experiment (BASJE) site at Mount Chacaltaya \cite{2014NIMPA.763..320T}, and the NICHE array at the TA Middle Drum site in Utah \cite{2019EPJWC.21005001B}. Both of these ground-based experiments use arrays of optical Cherenkov detectors to sufficiently sample the EAS Cherenkov light spatial and temporal profile to yield good measurements on the UHECR direction, energy, and particle type.
with high resolution.similar issues that are in inherent with single measurements of the optical Cherenkov radiation, e.g. a signal measurement does not precisely constrain the actual geometry of the observed EAS that is needed to measure the UHECR energy and location of \xmax~ with high resolution.

\section{UHECR EAS Radio Measurements using Balloon-borne Instruments}

\begin{figure}[t]
\begin{center}
{\includegraphics[width=0.68\textwidth]{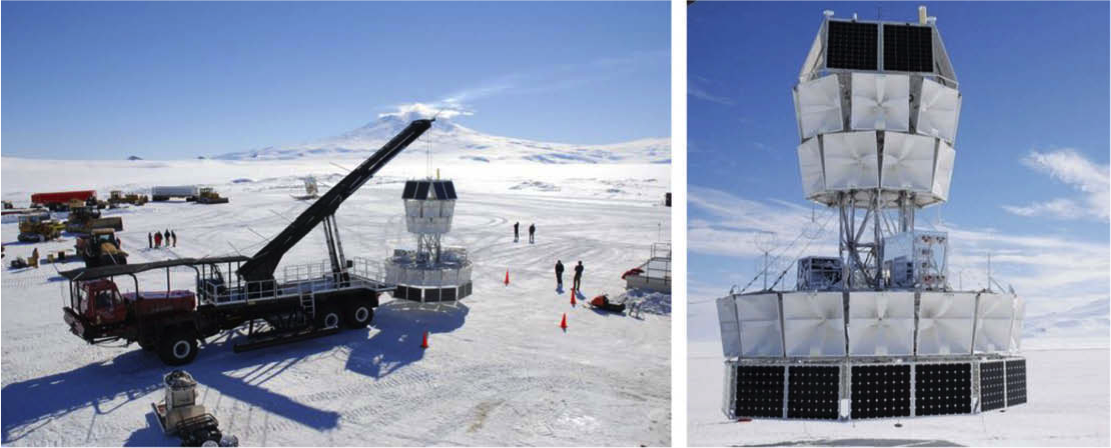}}
{\includegraphics[width=0.3\textwidth]{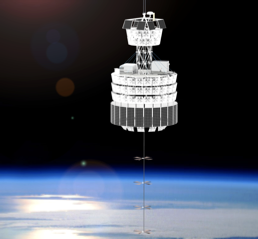}}
\caption{Leftmost: The ANITA-I long-duration balloon (LDB) payload preparing for launch in Antarctica. Middle: The ANITA-I instrument showing the upper 8-horn radio antennae array and lower 16-horn array, before its 35-day LDB flight in December 2006 through January 2007. From Ref. \citenum{2009APh....32...10A}.
  DOI: \href{https://doi.org/10.1016/j.astropartphys.2009.05.003}{10.1016/j.astropartphys.2009.05.003}.
  Rightmost: An artist rendering of the PUEO long-duration balloon payload to be launched in Antarctica, showing the various antenna  arrays including a low-frequency drop down antenna. From Ref. \citenum{2020arXiv201002892A}.
  DOI: \href{https://doi.org/10.1088/1748-0221/16/08/P08035}{ 	10.1088/1748-0221/16/08/P08035}.
  \label{PUEOpict}}
\end{center}
\end{figure}

To date, the most successful balloon-borne cosmic neutrino and in turn, UHECR measurements have been made using the EAS radio measurement technique pioneered by the Antarctic Impulsive Transient Antenna (ANITA) experiment. The left two picture in Fig>~\ref{PUEOpict} show the launch preparation and instrument from the ANITA-I flight. During
the four long-duration balloon (LDB) flights of the ANITA instrument \cite{2009APh....32...10A,2010PhRvD..82b2004G,2019EPJWC.21601009V,2019PhRvD..99l2001G}, the experiment has searched for neutrino-induced particle cascades, e.g. showers, in the Antarctic ice, beamed Askaryan radio emission \cite{Askaryan:1962hbi}, but has also has detected direct geomagnetic radio emission from UHECRs viewed above the Earth limb as well as the reflected UHECR EAS radio signal off of the ice \cite{Hoover_2010,2021PhRvL.126g1103G}. The  ANITA-IV radio measurements \cite{2019PhRvD..99l2001G, 2020arXiv200805690A} also included a class
of events consistent with being from Earth-emerging \taons observed near the Earth's limb from \Enu $\gsim$ 1 EeV. 
 The ANITA-IV results as include two events consistent with over-the-limb directly observed UHECRs events. The analysis included modeling the effects of atmospheric refraction for viewing just above the Earth limb events. Atmospheric refraction of optical and radio signals is a  potential VHECR/UHECR background to events from tau neutrino interacting in the Earth that lead to upward-moving EAS near the limb of the Earth  \cite{2020arXiv200805690A, 2019PhRvD..99l2001G, 2020PhRvD.102l3013V}.

The Payload for Ultrahigh Energy Observations (PUEO)  \cite{2020arXiv201002892A} is a NASA Pioneer-class mission for radio detection of neutrino events in the Antarctic ice,  \taon induced, upward-moving EAS from neutrino interactions in the Earth, and direct and reflected UHECR signals. The PUEO payload is shown in the rightmost panel in Fig.~\ref{PUEOpict} and is built on the ANITA development, flight, and analysis experience. PUEO employs a larger radio array in the 300$-$1200 MHz band, including a drop-down low-frequency, 50$-$300 MHz radio antenna. PUEO also includes an interferometric phased-array trigger, that coherently sums radio waveforms from each antennae with the appropriate propagation time delays to provide an increase in the signal-to-noise $\approx \sqrt{N_{\rm antennae}}$ compared to using a single antenna. This phased-array trigger technique has been demonstrated in the Askaryan Radio Array (ARA) experiment \cite{ALLISON2019112},

At frequencies $<10$~GHZ, well above the EAS radio detection frequency band, the atmosphere and clouds are radio transparent and do not affect the ability to observe radio emission of EAS. However, knowledge of the atmospheric properties is important for radio emission since the local index of refraction and the relative orientation of the geomagnetic field  determines the EAS radio beam pattern \cite{Alvarez-Muniz_2012}. However, similar issues that are in inherent with single measurements of the optical Cherenkov radiation, e.g. a signal measurement does not precisely constrain the actual geometry of the observed EAS that is needed to measure the UHECR energy and location of \xmax~ with high resolution.

Several studies of future space-based radio UHECR cascade detection, including the Earth-orbiting observatories, such as synoptic wideband orbiting radio detector (SWORD) to detect over-the-lib and reflected EAS signals \cite{Romero-Wolf_2013}, and lunar-orbiting missions, including the Zettavolt Askaryan Polarimeter (ZAP)  that detects cascades in the lunar regolith via the Askaryan effect \cite{2020arXiv200811232R}. One issue for Earth-orbiting experiments is  dealing with ionospheric dispersion of the EAS radio signal \cite{Romero-Wolf_2013}.

\section{Summmary}

Space-based UHECR missions provide extremely high UHECR exposure, $\sim 10^6$ km$^2$ ster years in a 5-year mission.  Using the stereo-fluorescence technique with at least two spacecraft also provide precision measurements to accurately determine the incident direction, energy, and nuclear composition of UHECRs.  Combined with full-sky sensitivity, space-based measurements provide a mechanism to study the astrophysics of UHECR sources.  While the majority of the development effort is to build and launch the spacecraft, once in orbit the operation can be relatively seamless and potentially longer than the designed lifetime, just dependent on expendables such as propulsion requirements for orbit maintenance, which is important for large, multi-meter sized spacecraft in LEO. Furthermore, the avionics (reaction wheels and magnetic torquer bars) required for satellite operation also allow for the spacecraft to be slewed and pointed to any direction along the orbit, with a minimal, if any, propulsion requirement.  The POEMMA spacecraft contain extra propulsion to allow the satellite separations to be changed multiple times per the mission \cite{2020PhRvD.102l3013V}.  Combined, the slewing and adjustment of the spacecraft separation allows tuning of the UHECR aperture. This if a recovery in the UHECR spectrum is observed above 100 EeV, the POEMMA telescope configuration could be optimized to increase the geometry factor at the highest energies, similar to that when POEMMA is in Earth-limb observing mode. The next generation of large ground-based UHECR experiments, such as the Global Cosmic Ray Observatory (GCOS) currently under study \cite{Hoerandel:2021qL} and the The Giant Radio Array for Neutrino Detection (GRAND) \cite{Fang_2017} if designed to cover an area of $\sim 4 \times 10^4$ km$^2$ would have slightly larger exposure to the space-based POEMMA experiment, but must be comprised of $>$ 10,000 individual surface detectors.  However, ground-detectors with the ability to discern electrons from muons as the EAS hits the Earth, especially combined with radio and air fluorescence measurements, would provide an unique measure of the hadronic interactions at the UHECR energy scale, which is well above that currently at terrestrial colliders, e.g. $\sqrt{s} = 450$ TeV for a 100 EeV proton. It should be noted that precision UHECR measurements can infer the proton-air cross-section at these energies since the UHECR composition, including proton content, is measured, e.g. see Ref.~\citenum{2020PhRvD.101b3012A}.

With the expected flight of EUSO-SPB2 in 2023, the technique of measuring VHECRs via the optical Cherenkov will be demonstrated while the flight of PUEO will make further EAS radio measurements on UHECRs.  This is a potential new technique for space-based VHECR and UHECR measurements, and could blossom with the use of a small constellation of SmallSats each with a modest-sized Cherenkov telescope. Combined with NASA Probe-class missions such as POEMMA, the future of space-based UHECR measurements looks very promising, and these high-statistics high-resolution measurements could finally provide an astrophysics accounting of the UHECR sources.

\bibliographystyle{ws-rv-van}
\bibliography{SpaceBasedUHECR}

\end{document}